\documentclass[10pt, draftclsnofoot, twocolumn]{IEEEtran}
\pdfoutput=1
\usepackage{amssymb}
\usepackage{algorithm}
\usepackage{algpseudocode}
\usepackage{algorithmicx}
\usepackage{amsfonts}
\usepackage{amsmath}
\usepackage{amssymb}
\usepackage{amsthm}
\usepackage{array}
\usepackage{bbding}
\usepackage{bigints}
\usepackage{booktabs}
\usepackage{cite}
\usepackage{geometry}
\usepackage{color}
\usepackage{diagbox}
\usepackage{epsfig,latexsym}
\usepackage{epstopdf}
\usepackage{graphicx}
\usepackage{fancyhdr}
\usepackage{float}
\usepackage{indentfirst}
\usepackage{lastpage}
\usepackage{makecell}
\usepackage{mathtools}
\usepackage{multirow}
\usepackage{pifont}
\usepackage{psfrag}
\usepackage{bm}
\usepackage{setspace}
\usepackage{stfloats}
\usepackage[colorlinks=true,linkcolor=black,citecolor=colorlinks=true,linkcolor=black,citecolor=blue,urlcolor=blue,urlcolor=blue]{hyperref}
\hypersetup{hidelinks,
	colorlinks=true,
	allcolors=black,
	pdfstartview=Fit,
	breaklinks=true}
\usepackage{cleveref}
\renewcommand{\baselinestretch}{1.0}
\usepackage{empheq}
\usepackage{enumerate}
\usepackage{times}
\usepackage{subfigure}
\renewcommand{\arraystretch}{1}
\newcolumntype{C}{>{\centering\arraybackslash$}p{\linewidth}<{$}}

\newtheorem{theorem}{Theorem}

\newtheorem{lemma}{Lemma}

\newtheorem{corollary}{Corollary}

\DeclareMathOperator*{\argmin}{arg\,min}
\DeclareMathOperator{\diag}{diag}
\DeclareMathOperator{\tr}{tr}

\DeclareMathOperator{\cov}{cov}
\DeclareMathOperator{\blkmtx}{blkmtx}

\newcommand{\RNum}[1]{\uppercase\expandafter{\romannumeral #1\relax}}

\newcounter{eqncnt}

\newcounter{eqnback}
\begin{document}

\title{MIMO-AFDM Outperforms MIMO-OFDM in the Face of Hardware Impairments}

\author{{Zeping Sui}, {\em Member,~IEEE}, {Zilong Liu}, {\em Senior Member,~IEEE}, {Leila Musavian}, {\em Senior Member,~IEEE}, {Yong Liang Guan}, {\em Senior Member,~IEEE}, {Lie-Liang Yang}, {\em Fellow,~IEEE}, and {Lajos Hanzo}, {\em Life Fellow,~IEEE}
\thanks{Zeping Sui, Zilong Liu, and Leila Musavian are with the School of Computer Science and Electronics Engineering, University of Essex, Colchester CO4 3SQ, U.K. (e-mail: zepingsui@outlook.com, \{zilong.liu,leila.musavian\}@essex.ac.uk).}
\thanks{Yong Liang Guan is with the School of Electrical and Electronic Engineering, Nanyang Technological University, 639798, Singapore (e-mail: eylguan@ntu.edu.sg).}
\thanks{Lie-Liang Yang and Lajos Hanzo are with the Department of Electronics and Computer Science, University of Southampton, Southampton SO17 1BJ, U.K. (e-mail: \{lly,lh\}@ecs.soton.ac.uk).}
}
\maketitle

\begin{abstract}
The impact of both multiplicative and additive hardware impairments (HWIs) on multiple-input multiple-output affine frequency division multiplexing (MIMO-AFDM) systems is investigated. For small-scale MIMO-AFDM systems, a tight bit error rate (BER) upper bound associated with the maximum likelihood (ML) detector is derived. By contrast, for large-scale systems, a closed-form BER approximation associated with the linear minimum mean squared error (LMMSE) detector is presented, including realistic imperfect channel estimation scenarios. Our first key observation is that the full diversity order of a hardware-impaired AFDM system remains unaffected, which is a unique advantage. Furthermore, our analysis shows that 1) the BER results derived accurately predict the simulated ML performance in moderate-to-high signal-to-noise ratios (SNRs), while the theoretical BER curve of the LMMSE detector closely matches that of the Monte-Carlo based one. 2) MIMO-AFDM is more resilient to multiplicative distortions, such as phase noise and carrier frequency offset, compared to its orthogonal frequency division multiplexing (OFDM) counterparts. This is attributed to its inherent chirp signal characteristics; 3) MIMO-AFDM consistently achieves superior BER performance compared to conventional MIMO-OFDM systems under the same additive HWI conditions, as well as different velocity values. The latter is because MIMO-AFDM is also resilient to the additional inter-carrier interference (ICI) imposed by the nonlinear distortions of additive HWIs. In a nutshell, compared to OFDM, AFDM demonstrates stronger ICI resilience and achieves the maximum full diversity attainable gain even under HWIs, thanks to its intrinsic chirp signalling structure as well as to the beneficial spreading effect of the discrete affine Fourier transform.
\end{abstract}
\begin{IEEEkeywords}
AFDM, hardware impairment, MIMO communications, OFDM, inter-carrier interference, performance analysis.
\end{IEEEkeywords}
\IEEEpeerreviewmaketitle

\section{Introduction}\label{Section 1}
\subsection{Background}
Next-generation wireless networks are expected to support efficient and reliable communications in the high-mobility environments of high-speed trains, vehicle-to-everything networks, flying drones, low-earth-orbit satellites, etc. However, high velocity typically results in severe inter-carrier interference (ICI) and inter-symbol interference (ISI) \cite{Wang20236G}. Hence, classic orthogonal frequency division multiplexing (OFDM) suffers from orthogonality erosion owing to high Doppler \cite{sui2025multi}. To address this challenge, both multicarrier orthogonal time frequency space (MC-OTFS) \cite{hadani2017orthogonal,10250854,10129061} and Zak-OTFS \cite{11083562} transceivers have been proposed. By sending the data symbols in the delay-Doppler domain, one is able to exploit the sparse and quasi-static wireless channels of OTFS systems. This leads to high Doppler resilience. However, as a two-dimensional (2D) modulation scheme, the transceiver complexity of OTFS is excessive, and it may require a significant change to the transceiver design compared to the family of OFDM systems. 

More recently, affine frequency division multiplexing (AFDM) was introduced in \cite{10087310}, whereby orthogonal chirp subcarriers are employed, and the information symbols are multiplexed in the discrete affine Fourier transform (DAFT)-domain. As a matter of fact, DAFT can be regarded as a generalization of both the discrete Fourier transform (DFT) and the discrete Fresnel Transform (DFnT) associated with the additional parameters $c_1$ and $c_2$, which are associated with the chirp rate and the initial phase, respectively. As a result, AFDM is highly flexible, subsumes both OFDM and orthogonal chirp division multiplexing (OCDM) as a pair of special cases, and it can be efficiently implemented through minimal modification of the OFDM architecture \cite{10087310}. Therefore, AFDM is recognized as a promising backward-compatible waveform that permits seamless integration into the existing OFDM-based wireless infrastructure\footnote{Let $M$ and $N$ denote the numbers of delay and Doppler bins in Zak-OTFS. Under the condition that a constant amplitude zero autocorrelation (CAZAC) basis is exploited to realize Zak-OTFS and the AFDM chirp parameters satisfy $c_1 MN, c_2 MN\in\mathbb{Z}$, Zak-OTFS and AFDM can be observed as unitarily equivalent waveforms \cite{11083562}.}. By appropriately tuning $c_1$ and $c_2$ based on the specific delay and Doppler profiles encountered, the non-zero elements of the effective DAFT-domain channel matrix associated with different paths can be well separated, thus yielding full diversity.
\subsection{Related Works}
\subsubsection{AFDM} {The bit error rate (BER) performance of single-input single-output (SISO)-AFDM using maximum likelihood (ML) detectors was derived in \cite{10087310} and extended to multiple-input multiple-output (MIMO)-AFDM systems in \cite{10557524}. It was also shown in \cite{10557524} that the sparse DAFT-domain channel matrix can be efficiently estimated by exploiting its diagonal structure. However, in large-scale AFDM systems having a high number of subcarriers, the complexity of the ML detector may become excessive. For low-complexity AFDM receiver design, the BER performance of a linear minimum mean squared error (LMMSE) detector was analyzed in \cite{10806672}. A pair of chirp parameter selection strategies were proposed for frequency selective and doubly selective channels, respectively. As a key enabler for high-mobility networks, AFDM has been combined with other emerging techniques, such as sparse code multiple access for supporting massive-scale connectivity \cite{10566604}, generalized spatial modulation for enhanced transmission reliability \cite{11185315}, bandwidth compression for higher spectral efficiency \cite{yi2025non}, and zero-padding for improved BER performance \cite{yi2025error}.}

{By exploiting the sparse structure of the DAFT-domain channel matrix, advanced signal processing algorithms, such as orthogonal approximate message passing \cite{10566604}, belief propagation \cite{11150613}, and expectation propagation \cite{10794592} were studied. To further improve the BER and spectral efficiency, index modulation aided AFDM systems were developed, where extra information bits are mapped onto the specific subcarrier activation patterns \cite{10342712,10845819}, and the chirp permutation domain \cite{10943004}.} 

{Another interesting application of AFDM is found in secure communications based on its unique features. Under the assumption of conventional OTFS and AFDM systems using brute-force demodulation, it was observed in \cite{11345476} that AFDM is more resilient to passive eavesdropping attacks than OTFS. Furthermore, by permuting the diagonal elements of the $c_2$-related chirp matrix, a secure AFDM system was designed in  \cite{11003930}.}

{Thanks to its unique chirp signalling property, AFDM can also be employed for the efficient design of multicarrier integrated sensing and communication systems \cite{11322564}. In short, AFDM is attractive due to its versatile capabilities as well as its backwards compatibility with OFDM \cite{10769778,rou2026afdm}, making it an excellent waveform candidate for next-generation networks.}

\subsubsection{Hardware Imperfections}
It is noted that the aforementioned AFDM studies were conducted under the assumption of either perfect hardware or perfect channel state information (CSI). In practical communication systems, severe performance loss may be encountered due to non-ideal hardware and/or imperfect CSI \cite{10841966,bjornson2015massive}. Some of the grave hardware impairments (HWIs), include oscillator phase noise (PN) \cite{10841966}, carrier frequency offset (CFO) \cite{10342857}, quantization distortions of low-resolution digital-to-analog converters (DACs) \cite{10041946}, in-phase and quadrature imbalance (IQI) \cite{7870582}, nonlinear power amplifier nonlinearity (NPA) \cite{5458079}, and direct current offset (DCO) \cite{yih2009analysis}. Physically, oscillator PN arises from thermal noise, flicker noise, and local oscillator (LO) imperfections, leading to random phase fluctuations that may destroy subcarrier orthogonality. CFO arises in realistic oscillators, which is caused by frequency mismatches between the transmit and receive LOs or by Doppler shifts, leading to a time-linear phase rotation that further impairs the received waveform.

As these front-end distortions accumulate, DAC limitations become another contributing factor. Specifically, DAC power consumption follows the relationship of $P_\text{DAC}\propto f_s\times 2^{b}$, where $b$ represents the quantizer resolution and $f_s$ denotes the sampling rate \cite{1532220}, making the design of high-resolution DACs challenging for high-throughput yet energy-efficient systems. When low-resolution DACs are employed, their coarse quantization introduces nonlinear distortions that further reduce the effective signal-to-noise ratio (SNR). Furthermore, from the mixer perspective, IQI induces additional distortions, such as mirror interference. This is due to amplitude and phase mismatches between the in-phase and quadrature branches, thereby deforming the received signal constellation \cite{4034110}. Furthermore, the above distortions are compounded when the power amplifier (PA) input signal enters the nonlinear region, where it imposes additional signal deformation and shortens the communication range. Finally, DCO arises from leakage, bias mismatches, and insufficient alternating current coupling, resulting in a static offset that limits the dynamic range.

Collectively, these HWIs originate from unavoidable imperfections in oscillators, converters, mixers, and PAs, and their combined effects can significantly degrade both the BER performance and SE. The above HWIs can be modelled as multiplicative and additive distortion terms that are uncorrelated with transmit/receive signals. These HWIs have been investigated in the OFDM literature. For example, the BER performance of sparse code multiple access-based OFDM systems impaired by CFO was analyzed in \cite{10342857}. The authors of \cite{yih2009analysis} analyzed the BER performance of OFDM under CFO as well as DCO, and proposed corresponding estimation and mitigation methods. Later in \cite{10746527}, low-complexity CFO estimation algorithms were proposed for OFDM systems, which can attain near-optimal performance. Considering a large-scale multiuser MIMO-OFDM system with low-resolution DACs, both BER performance and a lower bound of the sum rate were derived in \cite{jacobsson2019linear}. The authors of \cite{9399293} proposed an IQI estimation algorithm for wideband MIMO-OFDM systems, which also considered CFO. More recently, in \cite{11084986}, a meta-learning receiver was designed for a MIMO-OFDM system, whereby a deep neural network detector was proposed for nonlinear distortion compensation in the PA and denoising. However, the in-depth investigation of hardware-impaired AFDM systems is still an open issue.
\begin{table*}[t]
\footnotesize
\begin{center}
\caption{\textcolor{black}{Contrasting Contributions to the Related Hardware Impaired-OFDM/AFDM Literature.}}
\label{table1}
\begin{tabular}{l|c|c|c|c|c|c|c|c}
\hline
Contributions & \textbf{Our work} &  \cite{9562168} & \cite{10342857} & \cite{yih2009analysis} & \cite{10746527} & \cite{jacobsson2019linear} & \cite{9399293} & \cite{11084986} \\
\hline
\hline
Phase noise (PN) & \checkmark  & \checkmark &  &  & \checkmark &  & \\  
\hline
Carrier frequency offset (CFO) & \checkmark &  \checkmark   & \checkmark & \checkmark & & & \checkmark  \\ 
\hline
Low-resolution DACs & \checkmark &  &  &  &  & \checkmark & \\ 
\hline
In-phase and quadrature imbalance (IQI) & \checkmark  &   & & & &  & \checkmark &  \\ 
\hline
Nonlinear power amplifier (NPA) & \checkmark &     &  &  &  &  & & \checkmark \\ 
\hline
Direct current offset (DCO) & \checkmark    &  &  & \checkmark &  &  &  &  \\ 
\hline 
\textbf{BER upper bound of ML detector with HWIs} & \checkmark  &  &  &  & & & \\ 
\hline
\textbf{Approximated BER of LMMSE detector with HWIs} & \checkmark & & & &  &  &  \\ 
\hline
\textbf{Imperfect CSI and hardware} & \checkmark  &  & &  &   &  &   &  \\ 
\hline
\textbf{Diversity order analysis with HWIs} & \checkmark  &  & & & & & \\ 
\hline
\end{tabular}
\end{center}
\vspace{-3em}
\end{table*}
\subsection{Motivations and Contributions}
Against the above context, we aim to provide a comprehensive performance analysis of MIMO-AFDM systems, under both realistic HWIs and imperfect CSI, both quantitatively and qualitatively. Our research is driven by the following fundamental questions:
\begin{itemize}
\item How can the AFDM input-output relationship be modelled and modified when the aforementioned HWIs are incorporated?
\item How does the system perform under real-life practical scenarios? In particular, how to characterize AFDM's resilience to CFO and PN due to its chirp signalling properties?
\item How can the error performance of MIMO-AFDM be analytically characterized under HWIs and imperfect CSI across different system scales?
\end{itemize}

Our contributions are boldly contrasted to the existing AFDM and OFDM literature in Table \ref{table1}, which are detailed as follows:
\begin{itemize}
\item For the first time in the AFDM literature, we study the impact of various HWIs, such as PN, CFO, low-resolution DACs, IQI, NPA, and DCO. We first derive the DAFT-domain input-output relationship of MIMO-AFDM systems under realistic HWIs. Both multiplicative and additive distortion terms are modelled, yielding extra mirror interference, DC interference, and nonlinear-distortion interference terms.
\item Given a specific pairwise error event, we derive a closed-form expression of the BER upper bound associated with the ML detector. Moreover, we demonstrate that the channel diversity of MIMO-AFDM systems employing an ML detector remains full even under HWIs, since the multiplicative HWI-related matrices are diagonal. In addition, the signal-to-interference-plus-noise ratio (SINR) of each chirp subcarrier and the approximate BER of large-scale MIMO-AFDM under HWIs are derived using an LMMSE detector, whereby the channel estimation error is also considered.
\item Simulation results demonstrate that MIMO-AFDM is resilient to both PN and CFO, owing to its inherent chirp signal characteristics. By contrast, the BER performance of MIMO-OFDM becomes saturated in the high SNR region. Moreover, MIMO-AFDM consistently outperforms MIMO-OFDM under identical additive HWI settings. This is because MIMO-AFDM is capable of achieving stronger ICI resilience and full time-frequency diversity due to its inherent chirp characteristics and the spreading effect of DAFT modulation. Furthermore, the BER upper bound associated with the ML detector becomes tight as the SNR grows, while a close match is observed between the simulated BER using LMMSE detection and the approximate BER derived.
\end{itemize}
The rest of our paper is organized as follows. In Section \ref{Section 2}, we first present the conventional HWI-free AFDM system model, followed by a brief introduction to various HWIs in Section \ref{Section 3}. The end-to-end input-output relationship of MIMO-AFDM under realistic HWIs is derived in Section \ref{Section 4}, while we analyze the BER performance of ML and LMMSE detectors in both small- and large-scale MIMO-AFDM systems relying on both HWIs and imperfect CSI in Section \ref{Section 5}. Our simulation results are provided in Section \ref{Section 6}. Finally, we conclude in Section \ref{Section 7}.

\emph{Notation:} $\mathcal{CN}(\bm{a},\bm{B})$ represents a complex Gaussian distribution having mean vector $\bm{a}$ and covariance matrix $\bm{B}$. The vectors and matrices are denoted by lower- and upper-case boldface letters, respectively. The transpose, conjugate transpose and inverse of the matrix $\bm{A}$ are represented as $\bm{A}^T$, $\bm{A}^H$ and $\bm{A}^{-1}$, respectively. Furthermore, the delta function and uniform distribution between $a$ and $b$ are respectively denoted as $\delta(\cdot)$ and $\mathcal{U}[a,b]$. $\blkmtx\left(\{\bm{A}_{j,m}\}_{j=0,m=0}^{J-1,M-1}\right)$ denotes a block matrix with submatrices $\bm{A}_{j,m}$ in the $j$th row block and $m$th column block. The $l$th eigenvalue of $\bm{A}$ and the ring of integers are denoted by $\mu_l(\bm{A})$ and $\mathbb{Z}$, respectively. The real and imaginary parts of a complex number can be denoted by $\mathcal{R}\{\cdot\}$ and $\mathcal{I}\{\cdot\}$, respectively. The imaginary unit and the normalized discrete Fourier transform matrix are denoted as $\iota$ and $\bm{\mathcal{F}}$, respectively. The complementary error function is denoted by $erfc(\cdot)$. Finally, $\odot$ and $(\cdot)_N$ denote the Hadamard product and modulo-$N$ operators.

\section{HWI-free MIMO-AFDM System Model}\label{Section 2}
\subsection{AFDM Modulation}
\begin{figure*}[t]
\centering
\includegraphics[width=0.79\linewidth]{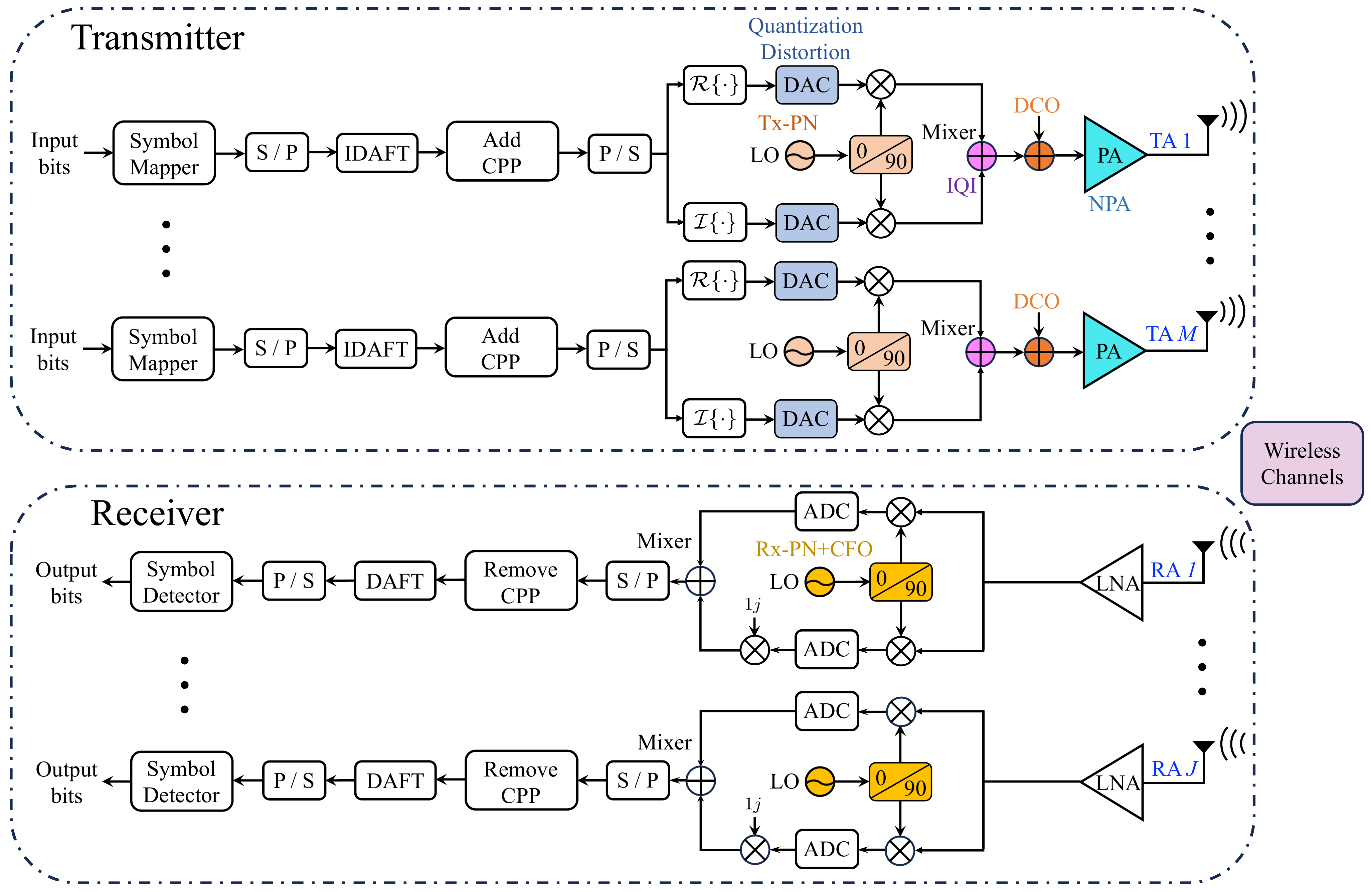}
\vspace{-1em}
\caption{{The block diagram and main circuits of MIMO-AFDM, where $M$ TAs and $J$ RAs are involved. LNA: low noise amplifier.}}
\label{Figure1}
\vspace{-2em}
\end{figure*}

Let us consider a limited-dimensional MIMO-AFDM system having $M$ transmit antennas (TAs) and $J$ receive antennas (RAs) relying on $N$ chirp subcarriers. We denote the subcarrier spacing and the symbol duration as $\Delta f$ and $T=1/\Delta f$, respectively. As seen in Fig. \ref{Figure1}, we can obtain the signal transmitted from the $m$th TA as $\bm{x}_m\in\mathcal{A}^{N}$ based on $|\mathcal{A}|$-ary QAM, where $|\mathcal{A}|$ represents the cardinality of the normalized signal constellation $\mathcal{A}$. Hence, the length of the transmit bit sequence is $L_b=MN\log_2 |\mathcal{A}|$. We can express the $m$th TD signal as
\begin{align}\label{eq_s}
	s_{m}(n)=\sum_{n'=0}^{N-1}x_{m}(n')\phi_n(n'),\quad n=0,\ldots,N-1,
\end{align}
where $\phi_n(n')=e^{\iota 2\pi[c_1 n^2+c_2 (n')^2+nn'/N]}/\sqrt{N}$ denotes the inverse DAFT kernel of the $n'$th chirp subcarrier parameterized by $c_1$ and $c_2$. It should be noted that the above model can be simplified to that of OFDM when $c_1=c_2=0$. We note that \eqref{eq_s} can be rewritten as 
\begin{align}\label{eq_s_vector}
\bm{s}_{m}=\bm{A}^H{\bm{x}}_{m}=\bm{\Lambda}^H_{c_1}\bm{\mathcal{F}}^H\bm{\Lambda}^H_{c_2}{\bm{x}}_{m},
\end{align}
where $\bm{A}=\bm{\Lambda}_{c_2}\bm{\mathcal{F}}\bm{\Lambda}_{c_1}$ denotes the DAFT matrix, and $\bm{\Lambda}_{c}=\diag\left(1,e^{-\iota2\pi c},\ldots,e^{-\iota 2\pi c(N-1)^2}\right)$ given the AFDM parameters $c\in\{c_1,c_2\}$. To harness circulant convolution and alleviate ISI, we employ a chirp-periodic prefix (CPP) of length $L_\text{CPP}$ within the AFDM frame, which can be formulated as
\begin{align}
	s_{m}(n)=s_m(N+n)e^{-\iota 2\pi c_1(N^2+2Nn)},	
\end{align}
for $n=-L_\text{CPP},\ldots,-1$.
\subsection{Channel Model}
{Given the $j$th TA and the $m$th RA for $j=0,\ldots,J-1$ and $m=0,\ldots,M-1$, let us consider that there are $P$ clusters of reflectors associated with the channel impulse response in the delay-Doppler domain, where $(h_{p,j,m},\tau_{p,j,m},\nu_{p,j,m})$ represent the channel gain, delay shifts, and Doppler shifts of the $p$th path from the TA $m$ to the RA $j$, respectively.} Hence, the channel impulse response can be formulated as
\begin{align}\label{eq_H}
	h_{j,m}(\tau,\nu)=\sum_{p=1}^{P} h_{p,j,m} \delta(\tau-\tau_{p,j,m})\delta(\nu-\nu_{p,j,m}),
	\end{align}
where $h_{p,j,m}\sim\mathcal{CN}(0,1/P)$. Given the bandwidth of $B=N\Delta f$, the sampling interval and the sampling rate are expressed as $T_s=1/B=1/(N\Delta f)$ and $f_s=1/T_s$, respectively. The Doppler and delay indices associated with the $p$th path can be respectively formulated as
\begin{align}
	\nu_{p,j,m}=\frac{k_{p,j,m}}{T}=\frac{k_{p,j,m} f_s}{N},\ \tau_p=\frac{l_{p,j,m}}{N\Delta f}={l_{p,j,m}}{T_s},
\end{align}
where $k_{p,j,m}=\alpha_{p,j,m}+\bar{\alpha}_{p,j,m}$ and $l_{p,j,m}$ represent the normalized Doppler and delay shifts, respectively. Furthermore, $\alpha_{p,j,m}\in[-k_{\text{max}},k_{\text{max}}]$ and $l_{p,j,m}\in[0,l_{\text{max}}]$ are the integer components, and the fractional Doppler shifts are denoted by $\bar{\alpha}_{p,j,m}\in[-1/2,1/2]$. The maximum delay is given by $\tau_\text{max}=l_\text{max}/(N\Delta f)$ with $l_\text{max}=L_P<N$. Upon sampling at $\{t=nT_s,n=0,\ldots,N-1\}$, the discrete delay-time (DT)-domain channel response between the $j$ th RA and $m$th TA can be formulated as
\begin{align}\label{eq_h_dis}
	\tilde{h}_{j,m}(n,u)=\sum_{p=1}^P h_{p,j,m}e^{j2\pi\frac{\nu_{p,j,m}}{f_s}(n-u)}\delta\left(u-\frac{\tau_{p,j,m}}{T_s}\right),
\end{align}
where $u=\tau/T_s$ denotes the normalized index of delay.
\subsection{AFDM Demodulation}
After discarding the CPP, the corresponding TD received signal between the $j$th RA and the $m$th TA can be formulated as
\begin{align}\label{eq_r_element}
	r_{j,m}(n)=\sum_{u=0}^{\infty}s_{m}(n-u)\tilde{h}_{j,m}(n,u)+\bar{w}_{j,m}(n),
\end{align}
for $n=0,\ldots,N-1$. Consequently, the TD signal received at the $j$th RA associated with $m$th TA can be formulated as $\bm{r}_{j,m}=\bar{\bm{H}}_{j,m}\bm{s}_{m}+\bar{\bm{w}}_{j,m}$, where $\bar{\bm{w}}_{j}$ is the TD additive white Gaussian noise (AWGN) vector, and the TD channel matrix can be expressed as \cite{10087310}
\begin{align}\label{eq_H_time}
	\bar{\bm{H}}_{j,m}=\sum_{p=1}^P h_{p,j,m}\bm{\Gamma}_{p,j,m}\bm{\Delta}_{k_{p,j,m}}\bm{\Pi}^{l_{p,j,m}},
\end{align}
where $\bm{\Pi}^{l_{p,j,m}}$ is the forward cyclic shift matrix associated with the delay, while the Doppler effect is characterized by $\bm{\Delta}_{k_{p,j,m}}=\diag\left\{1,e^{\iota 2\pi k_{p,j,m}/N},\ldots,e^{\iota 2\pi k_{p,j,m}(N-1)/N}\right\}$. Moreover, the CPP matrix is given by $\bm{\Gamma}_{p,j,m}=\diag\{\xi_{0,p,j,m},\ldots,\xi_{N-1,p,j,m}\}$ with
\begin{align}\label{eq_Gamma_element}
	\xi_{n,p,j,m}=\begin{cases}
		e^{-\iota 2\pi c_1 [N^2-2N(l_{p,j,m}-n)]}, & n<l_{p,j,m},\\
		1, & \text{otherwise}
			\end{cases}
\end{align}
for $n=0,\ldots,N-1$. Then, the DAFT-domain received signal can be formulated as
\begin{align}\label{eq_SISO}
	\bm{y}_{j,m}=\bm{A}\bm{r}_{j,m}=\bm{H}_{j,m}\bm{x}_{m}+\bm{w}_{j,m},
\end{align}
where the DAFT-domain channel matrix is given by
\begin{align}\label{eq_H_sub}
	\bm{H}_{j,m}=\bm{A}\bar{\bm{H}}_{j,m}\bm{A}^H.
\end{align}
\begin{figure}[t]
\centering
\includegraphics[width=0.5\linewidth]{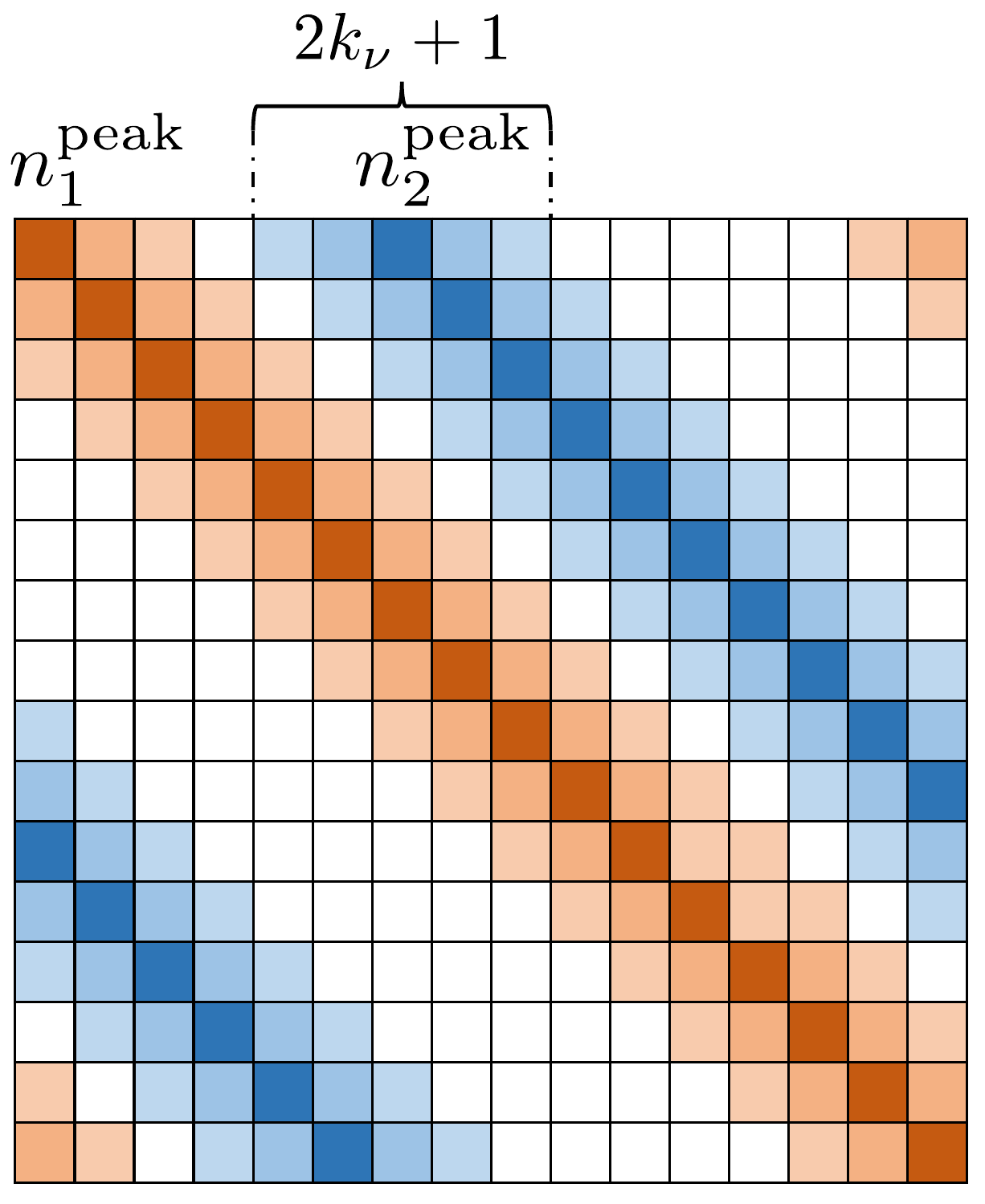}
\caption{The DAFT-domain channel matrix $\bm{H}_{j,m}$ with $N=16$, $k_{\nu}=2$, $l_\text{max}=1$, $k_\text{max}=1$, $\{k_1,k_2\}=\{0.3,1.2\}$ and $\{l_1,l_2\}=\{0,1\}$.}
\label{Figure2}
\vspace{1em}
\end{figure}
Based on \eqref{eq_H_time}, the DAFT-domain channel matrix can be rewritten as $\bm{H}_{j,m}=\sum_{p=1}^Ph_{p,j,m}\bm{H}_{p,j,m}$ with $\bm{H}_{p,j,m}=\bm{A}\bm{\Gamma}_{p,j,m}\bm{\Delta}_{k_{p,j,m}}\bm{\Pi}^{l_{p,j,m}}\bm{A}^H$. Consequently, the element-wise input-output relationship in the DAFT-domain can be formulated as
\begin{align}\label{eq_y_element}
	y_{j,m}(n)&=\frac{1}{N}\sum_{p=1}^P\sum_{n'=0}^{N-1}h_{p,j,m}\zeta_1(l_{p,j,m},n,n')\nonumber\\
	&\times\zeta_2(l_{p,j,m},k_{p,j,m},n,n')x_{m}(n')+w_{j}(n),
\end{align}
where we have
\begin{align}
	\zeta_1(l_{p,j,m},n,n')=e^{\iota \frac{2\pi}{N}\left[N c_1 l_{p,j,m}^2-n'l_{p,j,m}+Nc_2\left((n')^2-n^2\right)\right]},
\end{align}
\begin{align}
	\zeta_2(l_{p,j,m},k_{p,j,m},n,n')=\frac{e^{-\iota2\pi(n-n'+\text{Ind}_p)}-1}{e^{-\iota\frac{2\pi}{N}(n-n'+\text{Ind}_{p,j,m})}-1},
\end{align} 
with the index indicator $\text{Ind}_{p,j,m}=(k_{p,j,m}+2Nc_1 l_{p,j,m})_N$. Explicitly, each column and row of $\bm{H}_p$ only has $(2k_{\nu}+1)$ non-zero elements associated with the parameter $k_{\nu}$, and the indices of the central point of non-zero elements is given as $n^\text{peak}_p=\text{round}[(n+\text{Ind}_p)_N]$. Hence, \eqref{eq_y_element} can be rewritten as
\begin{align}
	y_{j,m}(n)\approx&\frac{1}{N}\sum_{p=1}^P\sum_{n'=(n^\text{peak}_p-k_\nu)_N}^{(n^\text{peak}_p+k_\nu)_N}h_{p,j,m}\zeta_1(l_{p,j,m},n,n')\nonumber\\
	&\times\zeta_2(l_{p,j,m},k_{p,j,m},n,n')x_{m}(n')+w_{j}(n).
\end{align}
Moreover, to achieve full channel diversity, the parameter $c_1$ is set as \cite{10087310}
\begin{align}\label{eq_c1}
	c_1=\frac{2(k_\text{max}+k_\nu)+1}{2N},
\end{align}
and the parameter $k_\nu$ should satisfy
\begin{align}\label{eq_N}
	2(k_\text{max}+k_\nu)(l_\text{max}+1)+l_\text{max}<N.
\end{align}
Under integer Doppler shift cases, we have $k_{\nu}=0$. The parameter $c_2$ can be set as a rational number sufficiently smaller than $1/(2N)$ or an arbitrary irrational number. Consequently, the non-zero elements of $\bm{H}_p$ with $\forall p\in\{1,\ldots,P\}$ can avoid overlapping. An example of the sparse structure of $\bm{H}_{j,m}$ is shown in Fig. \ref{Figure2}. Upon stacking the DAFT-domain transmit signal, the receive signal and AWGN vectors, we can obtain ${\bm{x}}=[{\bm{x}}_0^T,\ldots,{\bm{x}}_{M-1}^T]^T\in\mathbb{C}^{NM}$, $\bm{y}=[\bm{y}_0^T,\ldots,\bm{y}_{J-1}^T]^T\in\mathbb{C}^{NJ}$ and $\bm{w}=[\bm{w}_0^T,\ldots,\bm{w}_{J-1}^T]^T\in\mathbb{C}^{NJ}$. Furthermore, the DAFT-domain end-to-end MIMO channel matrix $\bm{H}\in\mathbb{C}^{NJ\times NM}$ can be formulated as
\begin{align}\label{eq_MIMO_channel}
	\bm{H}=\begin{bmatrix}
		\bm{H}_{0,0} & \cdots & \bm{H}_{0,M-1}\\
		\vdots & \ddots & \vdots\\
		\bm{H}_{J-1,0} & \cdots & \bm{H}_{J-1,M-1}			
		\end{bmatrix}.
\end{align}
Upon substituting \eqref{eq_H_time} and \eqref{eq_H_sub} into \eqref{eq_MIMO_channel}, we can rewrite \eqref{eq_MIMO_channel} as
\begin{align}\label{eq_H_final}
	\bm{H}=(\bm{I}_J\otimes\bm{A})\bar{\bm{H}}(\bm{I}_M\otimes\bm{A}^H),
\end{align}
where the $(j,m)$th component of $\bar{\bm{H}}$ can be attained based on \eqref{eq_H_time}. Finally, the end-to-end DAFT-domain input-output relationship can be formulated as
\begin{align}\label{eq_y}
	\bm{y}=\bm{H}{\bm{x}}+\bm{w},
\end{align}
where $\bm{w}\sim\mathcal{CN}(\bm{0},\sigma^2\bm{I}_{NJ})$ denotes the AWGN vector.

{In this Section, we have introduced the basis of MIMO-AFDM transceivers, where perfect hardware is considered. Next, in Section \ref{Section 3}, we will introduce various types of HWIs, and then formulate the end-to-end input-output relationship, when all the HWIs are characterized.}

\section{Hardware Impairments}\label{Section 3}
In this section, we introduce the multiplicative and additive distortions of various HWIs. Specifically, for our performance analysis, we assume having low-resolution DACs, IQI, DCO, and NPA at the transmitter side. The CFO appears at the receiver side, and both the transmitters and receivers are influenced by PN. {Moreover, except for PN, all types of HWIs are assumed to be identical across all antennas. The above HWIs and their impacts are summarized in Table \ref{table4}.}

\begin{table*}[t]
\centering
\footnotesize
\renewcommand{\arraystretch}{1.2} 
\caption{{Summary of HWIs and Their Impact on Receive Signals.}}
\label{table4}
\begin{tabular}{l|r|r|r}
\hline
\textbf{Type} & \textbf{HWI} & \textbf{Tx/Rx} & \textbf{Impact on receive signals} \\ \hline
\multirow{2}{*}{Multiplicative} 
& Phase noise (PN)  & Tx \& Rx & \multirow{2}{*}{\begin{tabular}[c]{@{}l@{}}Random phase rotation in DAFT-domain \end{tabular}} \\ \cline{2-3}
& Carrier frequency offset (CFO) & Rx        &  \\ \hline
\multirow{4}{*}{Additive} 
& Low-resolution DACs & Tx & \multirow{2}{*}{\begin{tabular}[c]{@{}l@{}}Reduced signal power \& distortion noise\end{tabular}} \\ \cline{2-3}
& Nonlinear power amplifier (NPA)                 & Tx &  \\ \cline{2-4}
& Direct current offset (DCO)                 & Tx & DC component as interference \\ \cline{2-4}
& In-phase and quadrature imbalance (IQI)                 & Tx & Mismatch between I- and Q- branches \\ \hline
\end{tabular}
\end{table*}

\subsection{Multiplicative Distortions}\label{Section 2-1}
\subsubsection{Phase noise}\label{Section 2-1-1}
Let us consider a practical scenario where the transceivers are equipped with free-running LOs, and the PN is generated within the baseband signal up-conversion and the passband signal down-conversion process. Explicitly, following the Wiener process, the time-domain (TD) PN terms associated with the transmitter (denoted by ``T'') and receiver (denoted by ``R'') sides can be respectively formulated as \cite{10841966,bjornson2015massive}
\begin{align}
	&\theta_{\text{T},m}(n)=\theta_{\text{T},m}(n-1)+\xi_{\text{T},m}(n),\xi_{\text{T},m}(n)\sim\mathcal{N}(0,\varsigma^2_{\text{T},m}), \nonumber\\
	&\theta_{\text{R},j}(n)=\theta_{\text{R},j}(n-1)+\xi_{\text{R},j}(n),\xi_{\text{R},j}(n)\sim\mathcal{N}(0,\varsigma^2_{\text{R},j}),
	\end{align}
for $n=1,\ldots,N-1$, where $n$ denotes the TD samples. Moreover, $\xi_{\text{T},m}(n)$ and $\xi_{\text{R},j}(n)$ denote the additive PN terms, while $\varsigma^2_{\text{T},m}$ and $\varsigma^2_{\text{R},j}$ are the corresponding PN variances, respectively. For our analysis, we assume $\varsigma^2_{\text{T},m}=\varsigma^2_{\text{T}}$ and $\varsigma^2_{\text{R},j}=\varsigma^2_{\text{R}}$, $\forall m,j$. Specifically, the variances can be formulated as $\varsigma^2_{i}=4\pi^2 f_c^2 \psi_i T_s$, where $\psi_i$ with $i\in\{\text{T},\text{R}\}$ representing the oscillator constants. Then, the $m$th transmit and $j$th receive PN diagonal matrices can be formulated as $\bm{\Phi}_{\text{T},m}=\diag\left\{e^{\iota\theta_{\text{T},m}(0)},\ldots,e^{\iota\theta_{\text{T},m}(N-1)}\right\}$ and $\bm{\Phi}_{\text{R},j}=\diag\left\{e^{\iota\theta_{\text{R},j}(0)},\ldots,e^{\iota\theta_{\text{R},j}(N-1)}\right\}$, respectively. Specifically, there are two types of transmit LOs \cite{bjornson2015massive}: 1) the transceivers are equipped with co-located arrays, yileidng a common LO (CLO) with identical $\bm{\Phi}_{\text{T},m}$ and $\bm{\Phi}_{\text{R},j}$, $\forall m,j$; 2) each TA and RA has distributed arrays relying on separate LOs (SLOs), hence the transmit PN components are independent. We consider both the cases of CLO and SLOs in this paper.
\subsubsection{Carrier frequency offset}\label{Section 3-1-2}
The CFO is imposed by the imperfect carrier synchronization at the receive LOs. The CFO matrix from the TA $m$ to the RA $j$ can be expressed as $\bm{P}_j=\diag\left\{1,e^{\iota2\pi\varphi_\text{CFO}/N},\ldots,e^{\iota2\pi\varphi_\text{CFO}(N-1)/N}\right\}$, where $\varphi_\text{CFO}$ denotes the normalized CFO factor \cite{10342857,9562168}.

Upon considering PN and CFO, the TD transmit and receive signals can be respectively formulated as
\begin{align}\label{eq_PN_CFO}
	\bm{s}_{m}\rightarrow\bm{\Phi}_{\text{T},m}\bm{s}_{m},\ \bm{r}_{j}\rightarrow\bm{P}_{j}\bm{\Phi}_{\text{R},j}\bm{r}_{j}.
\end{align}
\subsection{Additive Distortions}\label{Section 3-2}
\subsubsection{Low-resolution DACs}\label{Section 3-2-1}
\begin{table}[t]
\footnotesize
\begin{center}
\caption{DAC Scaling factors with $b\leq5$.}
\label{table2}
\begin{tabular}{c|ccccc}
\hline
$b$ & 1 & 2 & 3 & 4 & 5 \\
\hline
$\eta$ & 0.3634 & 0.1175 & 0.03454 & 0.009497 & 0.002499 \\
\hline
\end{tabular}
\end{center}
\end{table}
The additive quantization noise model (AQNM) is exploited to describe the output \cite{10041946}. Upon utilizing the Bussgang decomposition\footnote{Given two jointly circularly symmetric complex Gaussian random variables $x\in\mathbb{C}$ and $y\in\mathbb{C}$ and the deterministic function $z=U(x)\in\mathbb{C}$, then we have $\mathbb{E}\{zy^*\}=g\mathbb{E}\{xy^*\}$ with the Bussgang gain $g=\mathbb{E}\{U(x)x^*\}/\mathbb{E}\{|x|^2\}$. The deterministic function can be decomposed as $z=U(x)=gx+w$, where $w$ is a zero-mean complex random variable, which is uncorrelated with both $x$ and $y$.}, the output of DACs can be expressed as the scaled input signal plus uncorrelated noise, yielding 
\begin{align}\label{eq_quant}
	{\bm{s}}_m\rightarrow\mathbb{Q}(\bm{s}_m)=\sqrt{1-\eta}\bm{s}_m+\bm{n}_{m},
\end{align}
where $\eta\in[0,1]$ denotes the DAC scaling factor and $\bm{n}_{m}\sim\mathcal{CN}(\bm{0},\eta\bm{I}_N)$ is the quantization distortion noise vector. The DAC scaling factor $\eta$ depends on the number of (bits) $b$ used for quantization, whose values under $b\leq5$ are provided in Table \ref{table2}. By contrast, for $b>5$, the scaling factors can be formulated as $\eta\approx\sqrt{3}\pi2^{-2b-1}$ \cite{10041946}.
\subsubsection{In-phase and quadrature imbalance}\label{Section 3-2-2}
IQI is generated in the mixers of the direct conversion receivers. We assume that the receiver-side radio frequency (RF) chains are perfect, and we focus on the RF front-ends at the transmitter. Specifically, the IQI at the transmitter side occurs in TD, and the transmit vector can be rewritten as \cite{4034110}
\begin{align}
	\bm{s}_{m}\rightarrow\rho_1\bm{s}_{m}+\rho_2\bm{s}_{m}^*,
\end{align}
where the IQI parameters can be formulated as $\rho_1=\cos(\beta)+\iota\lambda\sin(\beta)$ and $\rho_2=\lambda\cos(\beta)-\iota\sin(\beta)$. Moreover, $\lambda\in[0,1]$ and $\beta\in[0,\frac{\pi}{2}]$ denote the gain and IQ-imbalanced phase between the sine and cosine components, respectively. Due to IQI, the received waveform contains a superposition of the desired signal and its conjugate component. In AFDM/OFDM systems, this implies that the observation at the $k$th subcarrier is contaminated by a conjugate ``mirror'' component originating from subcarrier $(N-k)$, resulting in ICI \cite{4034110}.
\subsubsection{Nonlinear power amplifier}\label{Section 3-2-3}
NPA refers to the nonlinear distortion imposed by the RF power amplifier when it operates in the saturated power region. Upon exploiting the Bussgang decomposition, the TD output of the nonlinear PA can be formulated as \cite{5458079}
\begin{align}
s_m(n)\rightarrow K_\text{PA} s_m(n)+q_m(n),
\end{align}
for $n=0,\ldots,N-1$ and $m=0,\ldots,M-1$, where $K_\text{PA}$ denotes the linear component gain, while $q_m(n)\sim\mathcal{CN}(0,\sigma^2_q)$ represents nonlinear distortion. Explicitly, we leverage the soft-envelope limiter (SEL) \cite{5458079} to describe the element-wise NPA, while $K_\text{PA}$ and $\sigma^2_q$ can be respectively formulated as
\begin{align}\label{eq_kpa}
	K_\text{PA}=1-\exp(-\nu_\text{clip}^2)+\frac{1}{2}\sqrt{\pi}\nu_\text{clip}erfc(\nu_\text{clip}),
\end{align}
\begin{align}\label{eq_sigma_q}
	\sigma^2_q=P_s\left[1-\exp(-\nu_\text{clip}^2)-K_\text{PA}^2\right].
\end{align}
In \eqref{eq_kpa} and \eqref{eq_sigma_q}, $\nu_\text{clip}=A_\text{is}/\sqrt{P_s}$ denotes the clipping level with average symbol power $P_s$ and input saturation voltage $A_\text{is}$.
\subsubsection{Direct current offset}\label{Section 3-2-4} The DCO includes self-mixing of LO leakage, transistor mismatches, and interference leakage \cite{yih2009analysis}. Consequently, the TD transmit signal can be modified as $\bm{s}_m\rightarrow\bm{s}_m+d_\text{T}\bm{1}_{N}$, where $d_\text{T}$ is the complex-valued DCO factor, and $\bm{1}_N$ is an all-one vector.

{The comprehensive types of HWIs have been introduced in this Section, then, we will work on system modelling upon considering the above HWIs.}

\section{Input-output Relationship with HWIs}\label{Section 4}
In this section, we derive the end-to-end input-output relationship of MIMO-AFDM systems by taking into account various HWIs introduced in Section \ref{Section 3}.

Upon considering all the above-mentioned transmit side HWIs, the TD transmit signal can be formulated as
\begin{align}\label{eq_check_s_m}
\check{\bm{s}}_m=K_\text{PA}\left(\rho_1\tilde{\bm{s}}_{\text{T},m}+\rho_2\tilde{\bm{s}}_{\text{T},m}^*+d_\text{T}\bm{1}_N\right)+\bm{q}_m,
\end{align}
where $\tilde{\bm{s}}_{\text{T},m}=\bm{\Phi}_{\text{T},m}\mathbb{Q}(\bm{s}_m)$. The power of the terms in the right-hand side of \eqref{eq_check_s_m} is given by $K_\text{PA}^2\rho_1^2N$, $K_\text{PA}^2\rho_2^2N$, $K_\text{PA}^2d_\text{T}^2N$, and $\sigma_q^2N$, respectively. The $(j,m)$th component of the end-to-end TD channel matrix ${\bm{\Theta}}$ is given by
\begin{align}\label{eq_Q_time_com}
	{\bm{\Theta}}_{j,m}=\sum_{p=1}^P{h}_{p,j,m}{\bm{B}}_{p,j,m},
\end{align}
where ${\bm{B}}_{p,j,m}=\bm{\Gamma}_{p,j,m}\bm{\Delta}_{k_{p,j,m}}\bm{\Pi}^{{l}_{p,j,m}}$. The TD input-output relationship is given by $\bar{\bm{r}}={\bm{\Theta}}\check{\bm{s}}+\tilde{\bm{w}}$, where $\check{\bm{s}}=[\check{\bm{s}}_0^T,\ldots,\check{\bm{s}}_{M-1}^T]^T$, $\tilde{\bm{w}}=[\bar{\bm{w}}_0^T,\ldots,\tilde{\bm{w}}_{J-1}^T]^T$ denotes the AWGN vector, and ${\bm{\Theta}}=\blkmtx\left(\{{\bm{\Theta}}_{j,m}\}_{j=0,m=0}^{J-1,M-1}\right)$. Then, based on \eqref{eq_PN_CFO}, the TD received signal involving CFO and reveiver-side PN can be expressed as $\tilde{\bm{r}}=\bm{P}\bm{\Phi}_\text{R}\bar{\bm{r}}$, where $\bm{P}=\diag\left\{\bm{P}_0,\ldots,\bm{P}_{J-1}\right\}\in\mathbb{C}^{NJ\times NJ}$ and $\bm{\Phi}_\text{R}=\diag\left\{\bm{\Phi}_{\text{R},0},\ldots,\bm{\Phi}_{\text{R},J-1}\right\}$ denote the block diagonal CFO and receive-PN matrices, respectively. Consequently, we can express the DAFT-domain received signal as
\begin{align}\label{eq_y_check}
	{\bm{y}}&=(\bm{I}_J\otimes\bm{A}) \bm{P}\bm{\Phi}_\text{R}{\bm{\Theta}}\check{\bm{s}}+(\bm{I}_J\otimes\bm{A}) \bm{P}\bm{\Phi}_\text{R}\tilde{\bm{w}},
\end{align} 
where ${\bm{y}}=\left[{\bm{y}}_0^T,\ldots,{\bm{y}}_{J-1}^T\right]^T$. Finally, let $\bm{\Phi}_\text{T}=\diag\left\{\bm{\Phi}_{\text{T},0},\ldots,\bm{\Phi}_{\text{T},M-1}\right\}$ and $\bm{q}=[\bm{q}_0^T,\ldots,\bm{q}_{M-1}^T]^T$. Then, based on \eqref{eq_quant}, \eqref{eq_check_s_m}, and \eqref{eq_y_check}, we can derive the end-to-end input-output relationship with HWI as \eqref{eq_HWI_IO}, which is shown at the top of the next page, and we have the following observations. {The main notations and definations of \eqref{eq_HWI_IO} are listed in Table \ref{Tab:Notations},} 
\setcounter{eqnback}{\value{equation}} \setcounter{equation}{30}
\begin{figure*}[!t]
\textcolor{black}{\begin{align}\label{eq_HWI_IO}
	{\bm{y}}=&\underbrace{\underbrace{\rho_1 K_\text{PA}\sqrt{1- \eta}(\bm{I}_J\otimes\bm{A}) \bm{P}\bm{\Phi}_\text{R}\bm{\Theta}\bm{\Phi}_\text{T}(\bm{I}_M\otimes\bm{A}^H)}_{{\bm{H}_\text{eff}}: \text{DAFT-domain effective channel matrix}}\bm{x}}_{\text{desired signal}}+\underbrace{\rho_2 K_\text{PA}\sqrt{1- \eta}(\bm{I}_J\otimes\bm{A}) \bm{P}\bm{\Phi}_\text{R}\bm{\Theta}\bm{\Phi}_\text{T}^*(\bm{I}_M\otimes\bm{A}^H)\bm{x}^*}_{\bm{v}_\text{MI}: \text{mirror interference}}\nonumber\\
	&+\underbrace{K_\text{PA}d_\text{T}(\bm{I}_J\otimes\bm{A})\bm{P}\bm{\Phi}_\text{R}\bm{\Theta} \bm{1}_{NM}}_{\bm{v}_\text{DI}: \text{DC interference}}+\underbrace{(\bm{I}_J\otimes\bm{A}) \bm{P}\bm{\Phi}_\text{R}\bm{\Theta}(K_\text{PA}\bm{\Phi}_\text{T}\bm{n}+\bm{q})}_{\bm{v}_\text{NI}: \text{nonlinear-distortion interference}}+\underbrace{(\bm{I}_J\otimes\bm{A}) \bm{P}\bm{\Phi}_\text{R}\tilde{\bm{w}}}_{\bar{\bm{w}}: \text{noise}},
\end{align}}%
\hrulefill
\vspace{-2em}
\end{figure*}
\setcounter{eqncnt}{\value{equation}}
\setcounter{equation}{\value{eqnback}}

\begin{table}[!t]
\caption{{List of Main Definitions in \eqref{eq_HWI_IO}}}
\vspace{-1em}
\label{Tab:Notations}
\begin{center}
\footnotesize
\begin{tabular}{l|r}
\hline
Notation & Definition \\
\hline
\hline
$M$ & Number of TAs \\
\hline
$J$ & Number of RAs \\
\hline
{$N$} & Number of subcarriers \\
\hline
$\rho_1,\ \rho_2$ & IQI parameters \\
\hline
$K_\text{PA}$ & Linear gain of PA \\
\hline
$\eta$ & DAC scaling factor \\
\hline
$\bm{A}$ & DAFT matrix \\
\hline
$\bm{P}$ & CFO matrix \\
\hline
$\bm{\Phi}_\text{T},\ \bm{\Phi}_\text{T}$ & Transmit and receive CFO matrices \\
\hline
$\bm{\Theta}$ & HWI-free TD channel matrix \\
\hline
$\bm{x}$ & 	DAFT-domain transmit symbol vector \\
\hline
$d_\text{T}$ & DCO factor \\
\hline
$\bm{n}$ & DAC quantization noise \\
\hline
$\bm{q}$ & Nonlienar distortion of PA \\
\hline
\end{tabular}
\end{center}
\end{table}

\begin{enumerate}
	\item The HWI-imposed DAFT-domain effective channel matrix in the first desired signal term can be regarded as the conventional HWI-free channel matrix of \eqref{eq_H_final} multiplied by HWI factors, diagonal PN matrices, and diagonal CFO matrices. Therefore, the number of non-zero elements of the effective hardware-impaired channel matrix ${\bm{H}}_\text{eff}$ retains the same sparse structure as that of its HWI-free counterpart. This property renders the family of sparse channel matrix-based channel estimation and data detection algorithms eminently applicable, such as message passing and sparse Bayesian learning.
	\item The second term $\bm{v}_\text{MI}$ of \eqref{eq_HWI_IO} is the mirror interference associated with all the HWIs and the conjugation of the transmit symbol vector. The power difference between the desired signal and mirror interference terms is only associated with $\rho_1$ and $\rho_2$. Moreover, the non-zero elements of the effective channel matrices in the desired signal and mirror interference terms are fully overlapped. Hence, the mirror interference term may result in significant SINR loss.
	\item The third term $\bm{v}_\text{DI}$ of \eqref{eq_HWI_IO} denotes the DC interference term imposed by the DCO, NPA, PN, and CFO, which can be considered as a DC component of the receive DAFT-domain signal.
	\item The fourth term $\bm{v}_\text{NI}$ of \eqref{eq_HWI_IO} represents the nonlinear-distortion interference term, which corresponds to the nonlinear-distortion noise of the NPA and low-resolution DACs, as well as transmit PN and CFO.
	\item The fifth term $\bar{\bm{w}}$ of \eqref{eq_HWI_IO} represents the Gaussian noise term associated with the received CFO and PN.
	\item Let us for a moment consider ideal hardware with $\eta=0$, $\bm{\Phi}_{\text{T}}=\bm{I}_{NM}$, $\bm{\Phi}_{\text{R}}=\bm{P}=\bm{I}_{NJ}$, $\rho_{1}=1$, $\rho_{2}=0$, $K_\text{PA}=1$, $\sigma^2_q=0$, and $d_{\text{T}}=0$. Then the DAFT-domain end-to-end input-output relationship of \eqref{eq_HWI_IO} can converge to that of the HWI-free system of \eqref{eq_y}.
	\item The covariance matrix of the Gaussian noise term $\bar{\bm{w}}$ of \eqref{eq_HWI_IO} can be expressed as $\cov(\bar{\bm{w}})=\sigma^2\bm{I}_{NJ}$.
	\item {We emphasize that different types of HWIs in \eqref{eq_HWI_IO} are independent, and the input-output relationship still holds when we remove any types of HWIs.}
\end{enumerate}

{Following our input-output relationship derived in \eqref{eq_HWI_IO}, we will carry out the analysis of error rate for both ML and LMMSE detectors in Section \ref{Section 5}.}

\section{Error Rate Analysis}\label{Section 5}
In this section, we analyze the BER performance of hardware-impaired MIMO-AFDM systems using the ML detector in small-scale systems and the LMMSE detectors for large-scale systems, respectively. Both ideal and imperfect CSI scenarios are considered.
\subsection{Error Performance Analysis using ML detector}\label{Section 5-1}
\subsubsection{Perfect CSI Conditions}\label{Section 5-1-1}
The signal received at the $j$th RA from the $m$th TA can be formulated based on \eqref{eq_Q_time_com} and \eqref{eq_HWI_IO} as
\begin{align}\setcounter{equation}{31}
\bm{y}_{j,m}=\bm{\Xi}_{j,m}(\bm{x}_m)\bm{h}_{j,m}+\bar{\bm{w}}_{j,m},
\end{align}
where $\bm{h}_{j,m}=[{h}_{1,j,m},\ldots,{h}_{P,j,m}]^T$. Let us denote the effective codeword matrix as $\bm{\Xi}_{j,m}(\bm{x}_m)=\left[\bm{\Xi}_{1,j,m}(\bm{x}_m),\ldots,\bm{\Xi}_{P,j,m}(\bm{x}_m)\right]\in\mathbb{C}^{N\times P}$ along with
\begin{align}
	\bm{\Xi}_{p,j,m}(\bm{x}_m)&=\rho_{1}\bm{\Upsilon}^1_{p,j,m}\bm{x}_m+\rho_{2}\tilde{\bm{\Upsilon}}^1_{p,j,m}\bm{x}_m^*\nonumber\\
	&+K_\text{PA}d_\text{T}\bm{\Upsilon}^2_{p,j,m}\bm{1}_N+\bm{\Upsilon}^2_{p,j,m}\bm{u}_{j,m},
\end{align}
where we leverage $\bm{u}_{j,m}=K_\text{PA}\bm{\Phi}_{\text{R},j}\bm{n}_{m}+\bm{q}_m$, and 
\begin{align}
	\bm{\Upsilon}^1_{p,j,m}&=K_\text{PA}\sqrt{1-\eta}\bm{\Upsilon}^2_{p,j,m}\bm{B}_{p,j,m}\bm{\Phi}_{\text{T},m}\bm{A}^H\nonumber\\
    \tilde{\bm{\Upsilon}}^1_{p,j,m}&=K_\text{PA}\sqrt{1-\eta}\bm{\Upsilon}^2_{p,j,m}\bm{B}_{p,j,m}\bm{\Phi}_{\text{T},m}^*\bm{A}^H,
	\end{align}
with $\bm{\Upsilon}^2_{p,j,m}=\bm{A}\bm{P}_j\bm{\Phi}_{\text{R},j}\bm{B}_{p,j,m}$ for $p=1,\ldots,P$. Let us furthermore represent the effective stacked channel matrices and the channel coefficient vector as $\bm{\Xi}_{j}(\bm{x})=[\bm{\Xi}_{j,0}(\bm{x}_0),\ldots,\bm{\Xi}_{j,M-1}(\bm{x}_{M-1})]\in\mathbb{C}^{N\times MP}$ and $\bm{h}_j=\left[\bm{h}_{j,0}^T,\ldots,\bm{h}_{j,M-1}^T\right]^T$, respectively. Hence, the $j$th received signal can be formulated as $\bm{y}_j=\sum_{m=0}^{M-1}\bm{y}_{jm}=\bm{\Xi}_{j}(\bm{x})\bm{h}_{j}+\bar{\bm{w}}_{j}$. The input-output relationship of \eqref{eq_HWI_IO} may then be rewritten as
\begin{align}\label{eq_IO_re}
	\bm{y}=\bm{\Xi}(\bm{x})\bm{h}+\bar{\bm{w}},
\end{align}
where $\bm{\Xi}(\bm{x})=\diag\{\bm{\Xi}_{0}(\bm{x}),\ldots,\bm{\Xi}_{J-1}(\bm{x})\}\in\mathbb{C}^{NJ\times PMJ}$ and $\bm{h}=[\bm{h}_0^T,\ldots,\bm{h}_{J-1}^T]^T\in\mathbb{C}^{PMJ}$. Based on \eqref{eq_IO_re}, the ML detector can be formulated as
\begin{align}
    \bm{x}^\text{ML}=\argmin_{\bm{x}\in\mathcal{A}^{NM}}\left \| \bm{y}-\bm{\Xi}(\bm{x})\bm{h} \right \|^2.
\end{align}
Let us consider the pairwise error event $\{\bm{x}^c,\bm{x}^e\}$, where $\bm{x}^c$ denotes the transmit symbol vector, while $\bm{x}^e$ represents the error-infested received symbol vector. By defining the error vector as $\bm{e}=\bm{x}^e-\bm{x}^c$, the effective Euclidean distance associated with this error event is given as
\begin{align}\label{eq_dE}
	d_E=||\bm{\Xi}(\bm{e})\bm{h}||^2=\bm{h}^H\left[\bm{\Xi}(\bm{e})\right]^H\bm{\Xi}(\bm{e})\bm{h}=\bm{h}^H\bm{\Omega}\bm{h},
\end{align}
where $\bm{\Xi}(\bm{e})=\bm{\Xi}(\bm{x}^e)-\bm{\Xi}(\bm{x}^c)$ denotes the codeword difference matrix, and the $L\times L$-dimensional codeword distance matrix associated with $L=PMJ$ can be formulated as $\bm{\Omega}=\diag\left\{\bm{\Omega}_0,\ldots,\bm{\Omega}_{J-1}\right\}$, where $\bm{\Omega}_j(\bm{e})=[\bm{\Xi}_{j}(\bm{e})]^H\bm{\Xi}_{j}(\bm{e})$ for $j=0,\ldots,J-1$. Hence, the CPEP can be formulated as
\begin{align}\label{eq_CPEP}
P(\bm{x}^c,\bm{x}^e|\bm{h})&=P\left[||\bm{y}-\bm{\Xi}(\bm{x}^e)\bm{h}||^2\le||\bm{y}-\bm{\Xi}(\bm{x}^c)\bm{h}||^2\right]\nonumber\\
&=P\left[2\Re\left\{\bm{y}^H\bm{\Xi}(\bm{x}^e)\bm{h}\right\}-2\Re\left\{\bm{y}^H\bm{\Xi}(\bm{x}^c)\bm{h}\right\}\right.\nonumber\\
&\left.\quad\geq||\bm{\Xi}(\bm{x}^e)\bm{h}||^2-||\bm{\Xi}(\bm{x}^c)\bm{h}||^2\right].
\end{align}
By using \eqref{eq_IO_re} and substituting the codeword difference matrix $\bm{\Xi}(\bm{x}^e)$ into \eqref{eq_CPEP}, we can further rewrite \eqref{eq_CPEP} as
\begin{align}\label{eq_CPEP2}
P(\bm{x}^c,\bm{x}^e|\bm{h})=P\left[\Re\{\bar{\bm{w}}^H\bm{\Xi}(\bm{e})\bm{h}\}\geq\frac{\left \|\bm{\Xi}(\bm{e})\bm{h}\right \|^2}{2}\right].
\end{align}
It can be readily shown that $\Re\{\bar{\bm{w}}^H\bm{\Xi}(\bm{e})\bm{h}\}$ follows the Gaussian distribution with zero mean and a variance of ${\sigma^2 d^2_E}/{2}$. Hence, we have
\begin{align}\label{eq_CPEP3}
P(\bm{x}^c,\bm{x}^e|\bm{h})=Q\left(\sqrt{\frac{d_E}{2\sigma^2}}\right),
\end{align}
where $Q(\cdot)$ denotes the Gaussian $Q$-function of $Q(x)\approx\frac{1}{12}\exp(-{x^2}/{2})+\frac{1}{4}\exp(-{2x^2}/{3})$ \cite{1188428}. Hence, the unconditional PEP (UPEP) can be expressed based on \eqref{eq_CPEP3} as
\begin{align}\label{eq_UPEP1}
	P(\bm{x}^c,\bm{x}^e)=\mathbb{E}_{\bm{h}}\left\{\frac{1}{12}\exp(-\gamma_1 d_E)+\frac{1}{4}\exp(-\gamma_2 d_E)\right\},
\end{align}
where $\gamma_1=1/(4\sigma^2)$ and $\gamma_2=1/(3\sigma^2)$. Let $r=\text{rank}(\bm{\Gamma})$, where $\bm{\Gamma}=\mathbb{E}\{\bm{h}\bm{h}^H\}$ is the channel coefficient covariance matrix. Then, we can derive $\bm{\Gamma}=\bm{Z}\bm{U}\bm{Z}^H$ and $\bm{h}=\bm{Z}\bm{u}$ with a diagonal $\mathbb{E}\{\bm{u}\bm{u}^H\}=\bm{U}\in\mathbb{C}^{r\times r}$ \cite{johnsoncambridge}. Hence, the probability density function (PDF) of $\bm{u}$ is given by
\begin{align}\label{eq_PDF_u}
	f(\bm{u})=\frac{\pi^{-r}}{\det(\bm{U})}\exp(-\bm{u}^H\bm{U}^{-1}\bm{u}).
\end{align}
Using \eqref{eq_PDF_u} to calculate the expectation operation of \eqref{eq_UPEP1} and following some algebraic simplifications, the UPEP can be derived as
\begin{align}\label{eq_UPEP2}
	P(\bm{x}^c,\bm{x}^e)\approx\frac{1}{12\det(\bm{I}_{L}+\gamma_1\bm{\Gamma}\bm{\Omega})}+\frac{1}{4\det(\bm{I}_{L}+\gamma_2\bm{\Gamma}\bm{\Omega})}.
\end{align}
Let us now dfine $\bm{D}_i=\bm{I}_{L}+\gamma_i\bm{K}$ with $\bm{K}=\bm{\Gamma}\bm{\Omega}$ for $i=1,2$. We can readily show that $\det(\bm{D}_i)=\prod_{l=1}^L\mu_l(\bm{D}_i)=\prod_{l=1}^R[1+\gamma_i\mu_l(\bm{K})]$, where $R=\text{rank}(\bm{K})$. In the high-SNR region of $\gamma_i\gg1$, \eqref{eq_UPEP2} can be rewritten as
\begin{align}\label{eq_UPEP3}
P(\bm{x}^c,\bm{x}^e)\approx\left[12\gamma_1^R\prod_{l=1}^R\mu_l(\bm{K})\right]^{-1}+\left[4\gamma_2^R\prod_{l=1}^R\mu_l(\bm{K})\right]^{-1}.
\end{align}

Finally, upon exploiting the union bounding technique, the average BER (ABER) of the hardware-impaired MIMO-AFDM system can be approximated by
\begin{align}\label{eq_ABER}
{P}_a\approx\frac{1}{2^{L_b}L_b}\sum_{\bm{b}^c}\sum_{\bm{b}^e}D(\bm{b}^e,\bm{b}^c)P(\bm{x}^c,\bm{x}^e),
\end{align}
where $D(\bm{b}^e,\bm{b}^c)$ represents the Hamming distance between the bit sequences $\bm{b}^e$ and $\bm{b}^c$ representing the symbol vectors $\bm{x}^e$ and $\bm{x}^c$.

Next, let us analyze the system's diversity order and coding gain. Firstly, we can observe from \eqref{eq_UPEP3} that the diversity order can be formulated as $V_D=\min_{\forall \bm{e}}\{r,\text{rank}(\bm{\Omega})\}$. Explicitly, we have $r=PJ$, since the channel coefficients are generated independently. Moreover, it can be readily shown that the PN and CFO matrices $\bm{P}_j$, $\bm{\Phi}_{\text{T},m}$ and $\bm{\Phi}_{\text{R},j}$ are all {diagonal} matrices, $\forall j,m$, hence $\text{rank}(\bm{\Omega})$ is independent of the CFO and PN matrices. It has also been observed in \cite{10557524} that we have $\min_{\forall \bm{e},j}\text{rank}(\bm{\Omega}_j)=P,\forall j$. Furthermore, since the transmit symbols associated with each TA are independent, we have $\text{rank}(\bm{\Omega})=PJ$, i.e., the diversity order of our hardware-impaired MIMO-AFDM systems is $V_D=PJ$\footnote{{The diversity order characterizes the system's capability to combat channel fading \cite{1197843}, which can be reflected by the {BER curve slope} as a function of the SNR \cite{10087310,10557524,10342712}. By contrast, the degrees of freedom (DoF) are determined by the number of parallel independent information streams that can be transmitted at high SNRs, which is related to the {channel capacity}. Moreover, there is a fundamental trade-off between diversity and DoF as shown in \cite{1197843}.}}.
\subsubsection{Imperfect CSI Conditions}\label{Section 5-1-2}
Under realistic imperfect channel estimation scenarios, the channel coefficient vector can be expressed as $\check{\bm{h}}=\bm{h}+\bm{e}_h$ \cite{752125}, where the channel estimation error obeys $\bm{e}_h\sim\mathcal{CN}(\bm{0},\sigma_h^2\bm{I}_{L})$ with the variance $0\leq \sigma_h^2\leq 1$. Consequently, we can rewrite the input-output relationship of \eqref{eq_IO_re} as
\begin{align}
	\bm{y}=\bm{\Xi}(\bm{x})\check{\bm{h}}-\bm{\Xi}(\bm{x})\bm{e}_h+\bar{\bm{w}}=\bm{\Xi}(\bm{x})\check{\bm{h}}+\check{\bm{w}},
\end{align}
where $\check{\bm{w}}=-\bm{\Xi}(\bm{x})\bm{e}_h+\bar{\bm{w}}$ represents the effective noise term. Similar to \eqref{eq_CPEP2}, the CPEP relying on imperfect CSI can be formulated as
\begin{align}\label{eq_CPEPI1}
	P(\bm{x}^c,\bm{x}^e|\check{\bm{h}})=P\left[\Re\{\check{\bm{w}}^H\bm{\Xi}(\bm{e})\check{\bm{h}}\}\geq\frac{\left \|\bm{\Xi}(\bm{e})\check{\bm{h}}\right \|^2}{2}\right].
\end{align}
We can readily show that $\Re\{\check{\bm{w}}^H\bm{\Xi}(\bm{e})\check{\bm{h}}\}$ obeys a Gaussian distribution with a zero mean and a variance of
\begin{align}
	\sigma^2_I=\frac{\sigma_h^2\left \|\bm{\Xi}(\bm{x}^c)^H\bm{\Xi}(\bm{e})\check{\bm{h}}\right \|^2+\sigma^2\left \|\bm{\Xi}(\bm{e})\check{\bm{h}}\right \|^2}{2}.
\end{align}
Hence, the CPEP can be expressed as
\begin{align}\label{eq_CPEPI2}
	&P(\bm{x}^c,\bm{x}^e|\check{\bm{h}})\nonumber\\
	&=Q\left(\frac{\left \|\bm{\Xi}(\bm{e})\check{\bm{h}}\right \|^2}{\sqrt{2\sigma_h^2\left \|\bm{\Xi}(\bm{x}^c)^H\bm{\Xi}(\bm{e})\check{\bm{h}}\right \|^2+2\sigma^2\left \|\bm{\Xi}(\bm{e})\check{\bm{h}}\right \|^2}}\right).
\end{align}
Specifically, the UPEP can be formulated as $P(\bm{x}^c,\bm{x}^e)=\mathbb{E}_{\check{\bm{h}}}\left\{P(\bm{x}^c,\bm{x}^e|\check{\bm{h}})\right\}$, and $\check{\bm{h}}$ has the multivariate complex Gaussian PDF of $f(\check{\bm{h}})=\frac{\pi^{-L}}{\det({\bm{\Gamma}})}\exp\left(-\check{\bm{h}}^H{\check{\bm{\Gamma}}}^{-1}\check{\bm{h}}\right)$, where $\check{\bm{\Gamma}}=\mathbb{E}\left\{\check{\bm{h}}\check{\bm{h}}^H\right\}={\bm{\Gamma}}+\sigma_h^2\bm{I}_{L}$. Nonetheless, given the complex structure of \eqref{eq_CPEPI2}, it remains an open challenge how to directly calculate the UPEP. Since we leverage the normalized constellation $\mathcal{A}$ having a constant envelope, we have $\left\|\bm{\Xi}(\bm{x}^c)^H\bm{\Xi}(\bm{e})\check{\bm{h}}\right\|^2\leq\left\|\bm{\Xi}(\bm{e})\check{\bm{h}}\right \|^2$. Then, the CPEP upper-bound is given by
\begin{align}\label{eq_CPEPI3}
	P(\bm{x}^c,\bm{x}^e|\check{\bm{h}})=Q\left(\sqrt\frac{\left \|\bm{\Xi}(\bm{e})\check{\bm{h}}\right \|^2}{{2\sigma_h^2+2\sigma^2}}\right).
\end{align}
Then, the UPEP in the presence of channel estimation imperfections can be formulated similarly to \eqref{eq_UPEP2} as
\begin{align}\label{EqA11}
	&P(\bm{x}^c,\bm{x}^e)\approx\frac{1}{12\det(\bm{I}_{L}+\kappa_1\check{\bm{\Gamma}}\bm{\Omega})}+\frac{1}{4\det(\bm{I}_{L}+\kappa_2\check{\bm{\Gamma}}\bm{\Omega})},
\end{align}
where $\kappa_1=1/4(\sigma^2_h+\sigma^2)$ and $\kappa_2=1/3(\sigma^2_h+\sigma^2)$. Finally, the BER upper bound of imperfect CSI can be obtained based on \eqref{eq_ABER}.
\vspace{-1em}
\subsection{Performance Analysis of LMMSE Detection}\label{Section 5-2}
In this subsection, we leverage the assumption in \cite{7870582} that the mirror interference term $\bm{v}_\text{MI}$ represents interference. Upon considering imperfect CSI scenarios, the channel estimation error matrix can be expressed as $\tilde{\bm{H}}_\text{eff}={\bm{H}}_\text{eff}-\hat{\bm{H}}_\text{eff}$, where $\hat{\bm{H}}_\text{eff}$ denotes the estimated channel matrix. Moreover, the structure of the estimation error matrix $\tilde{\bm{H}}_\text{eff}$ is the same as ${\bm{H}}_\text{eff}$, whose non-zero elements obey $\mathcal{CN}(0,\sigma^2_h)$. Then, based on \eqref{eq_HWI_IO}, the end-to-end input-output relationship can be rewritten as
\begin{align}
	\bm{y}=\hat{\bm{H}}_\text{eff}\bm{x}+\tilde{\bm{H}}_\text{eff}\bm{x}+\bm{v},
\end{align}
where $\bm{v}=\bm{v}_\text{MI}+\bm{v}_\text{DI}+\bm{v}_\text{NI}+\bar{\bm{w}}$ represents the effective DAFT-domain noise term, whose covariance matrix $\bm{R}_{{v}}=\mathbb{E}\{\bm{v}\bm{v}^H\}$ can be calculated based on the popular Monte-Carlo averaging method. Upon exploiting the LMMSE detector, the estimated symbol vector can be expressed as
\begin{align}\label{eq_hat_x}
	\hat{\bm{x}}=\bm{G}\bm{y}&=\bm{G}(\hat{\bm{H}}_\text{eff}\bm{x}+\tilde{\bm{H}}_\text{eff}\bm{x}+\bm{v}),
\end{align}
where the LMMSE equalization matrix is $\bm{G}=\hat{\bm{H}}^H_\text{eff}\bm{W}^{-1}$, associated with the Hermitan matrix $\bm{W}=\hat{\bm{H}}_\text{eff}\hat{\bm{H}}_\text{eff}^H+\tilde{\bm{H}}_\text{eff}\tilde{\bm{H}}_\text{eff}^H+\bm{R}_v$. Therefore, we can write $\hat{\bm{H}}_\text{eff}=\bm{W}\bm{G}^H$. Let us denote the estimates of the $n$th symbol transmitted by the $m$th TA as $\hat{x}(c)$, where $c=Nm+n$ for $m=0,\ldots,M-1$ and $n=0,\ldots,N-1$. Hence, $\hat{x}(c)$ can be expressed as
\begin{align}
	\hat{x}(c)=T(c,c)x(c)+\sum_{d\neq c}T(c,d)x(d)+\tilde{v}(c),
\end{align}
where $\bm{T}=\bm{G}\hat{\bm{H}}_\text{eff}$ and $\tilde{\bm{v}}=\bm{G}\tilde{\bm{H}}_\text{eff}\bm{x}+\bm{G}\bm{v}$. The corresponding output SINR of the $c$th symbol can be formulated as
\begin{align}\label{eq_SINR_MMSE}
	\chi_c=\frac{\mathbb{E}\{|T(c,c)|^2\}}{\sum\limits_{d\neq c}\mathbb{E}\{|T(c,d)|^2\}+\mathbb{E}\{|\tilde{v}(c)|^2\}}.
\end{align}
According to \eqref{eq_hat_x}, the corvariance matrix of $\hat{\bm{x}}$ can be formulated as $\mathbb{E}\left\{\hat{\bm{x}}\hat{\bm{x}}^H\right\}=\bm{G}\bm{W}\bm{G}^H\overset{(a)}{=}\bm{G}\hat{\bm{H}}_\text{eff}=\bm{T}$, where $\overset{(a)}{=}$ exploits the fact that $\hat{\bm{H}}_\text{eff}=\bm{W}\bm{G}^H$. Hence, we have $\mathbb{E}\{|\hat{x}(c)|^2\}=T(c,c)$ and the denominator of \eqref{eq_SINR_MMSE} can be written as $\mathbb{E}\{|\hat{x}(c)|^2\}-|T(c,c)|^2=T(c,c)-|T(c,c)|^2$. Then, the SINR of \eqref{eq_SINR_MMSE} can be reformulated as
\begin{align}\label{eq_SINR}
	\chi_c=\frac{|T(c,c)|^2}{T(c,c)-|T(c,c)|^2}=\frac{T(c,c)}{1-T(c,c)}.
\end{align}
Then, upon using an $|\mathcal{A}|$-ary QAM constellation, the approximated BER can be formulated as \cite{proakis2001digital}
\begin{align}\label{eq_Pe}
	P_e&\approx\frac{1}{NM}\sum_{c=0}^{NM-1}u_1 Q\left(\sqrt{u_2\chi_c}\right),\nonumber\\
	u_1&=\frac{4}{\log_2 |\mathcal{A}|}\left(1-\frac{1}{\sqrt{|\mathcal{A}|}}\right),\ u_2=\frac{3}{|\mathcal{A}|-1},
\end{align}
where $|\mathcal{A}|$ denotes the modulation order. Similar to \cite{10806672}, we can readily show that the function $Q(\sqrt{u_2 x})$ is convex under each constellation mapping scheme. Finally, upon substituting \eqref{eq_SINR} into \eqref{eq_Pe} and harnessing Jensen's inequality, the BER lower bound can be formulated as
\begin{align}
	P_e\geq P_{l}=u_1 Q\left(\sqrt{\frac{u_2\tr(\bm{T})/(NM)}{1-\tr(\bm{T})/(NM)}}\right),
\end{align}
and the lower bound can be achieved when $T(c,c)={\tr(\bm{T})}/{NM},\ \forall c$, which implies that all the diagonal elements of $\bm{T}$ have the same value.
\section{Simulation Results}\label{Section 6}
In this section, we evaluate the overall system performance of MIMO-AFDM systems under realistic HWIs. We first characterize the BER performance of AFDM systems using an LMMSE detector under different HWIs, and evaluate the accuracy of our analytical results. {Moreover, we adopt the assumption in  \cite{10557524} that we have $\tau_{p,j,m}=\tau_p$ and $\nu_{p,j,m}=\nu_p$, $\forall j,m$.} Unless stated otherwise, {we adopt the parameters shown in Table \ref{Tab:Parameters}.} The $p$th normalized Doppler index is given by $k_p=k_{\max}\cos(\theta_p)$, with $\theta\in\mathcal{U}[0,\pi]$ based on Jakes' spectrum. We set $l_1=0$ and the normalized maximum delay shift as $l_{\max}=P-1$. Then, the delay indices associated with the $p$th path can be denoted as $l_p\in\mathcal{U}[1,l_{\max}]$, $\forall p\neq 1$. The $p$th complex-valued channel gain obeys $h_p\sim\mathcal{CN}(0,1/P)$.

\begin{table}[!t]
\caption{{Simulation Parameters}}
\vspace{-1em}
\label{Tab:Parameters}
\begin{center}
\footnotesize
\begin{tabular}{l|r}
\hline
Parameters & Values \\
\hline
\hline
Number of TAs, $M$ & 4 \\
\hline
Number of RAs, $J$ & 4 \\
\hline
Number of subcarriers, $N$ & 64 \\
\hline
Subcarrier spacing, $\Delta f$ & 15 kHz \\
\hline
Carrier frequency, $f_c$ & 4 GHz \\
\hline
Number of channel paths, $P$ & 3 \\
\hline
Velocity, $v$ & 540 km/h \\
\hline
Modulation type & QPSK \\
\hline
\end{tabular}
\end{center}
\end{table}

\begin{figure}[b]
\centering
\includegraphics[width=0.9\linewidth]{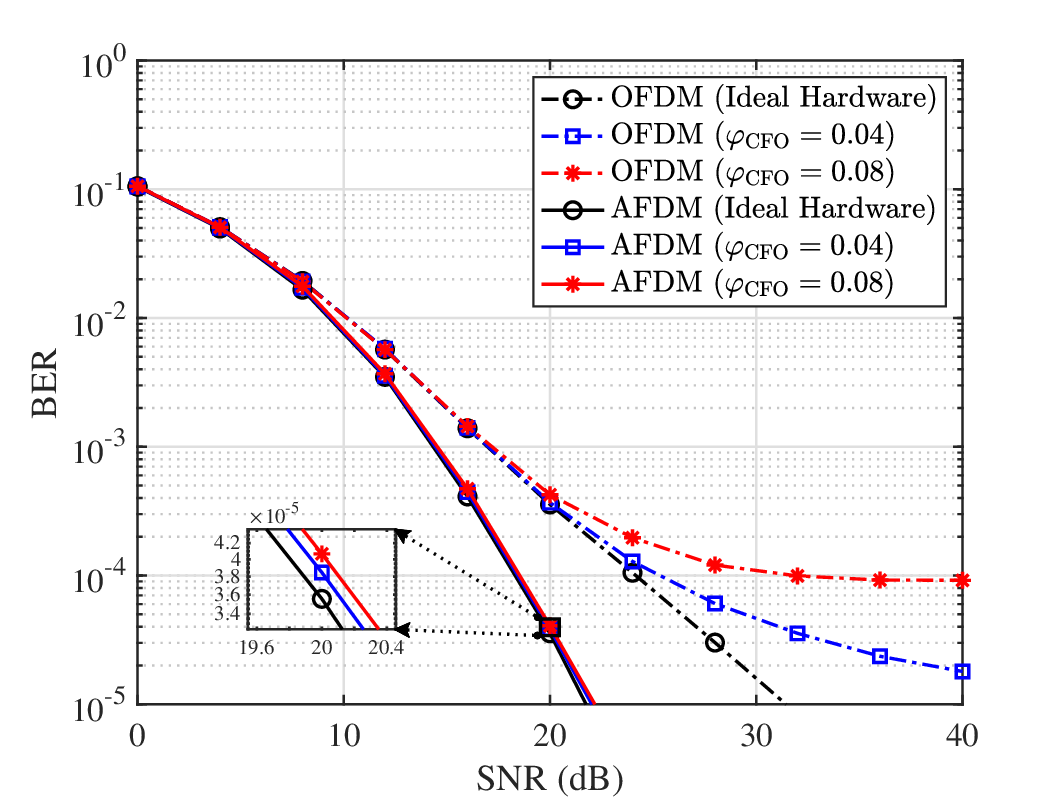}
\caption{BER performance of MIMO-AFDM/OFDM systems with different settings of normalized CFOs.}
\label{Figure3}
\end{figure}

In Fig. \ref{Figure3}, we investigate the BER performance of both MIMO-AFDM and MIMO-OFDM under different normalized CFOs. We can infer the following observations from Fig. \ref{Figure3}. Firstly, under ideal hardware, MIMO-AFDM achieves about $9$ dB SNR gain over MIMO-OFDM at the BER of $10^{-5}$ under high-mobility scenarios. Secondly, the BER performance of MIMO-AFDM impaired by CFOs remains nearly identical to that of the ideal hardware case. This trend is primarily due to AFDM's orthogonal chirp subcarriers, which have strong intrinsic resilience to both Doppler shifts and CFO. By contrast, at the BER of $10^{-4}$, MIMO-OFDM associated with $\varphi_\text{CFO}=0.08$ suffers an SNR loss of $8$ dB and $1$ dB for $\varphi_\text{CFO}=0.04$, respectively compared to the ideal hardware scenario. Moreover, the BER of MIMO-OFDM at $\varphi_\text{CFO}=0.08$ exhibits an error floor as the SNR increases, since CFO imposes significant ICI on the OFDM subcarriers.

\begin{figure}[t]
\centering
\includegraphics[width=0.9\linewidth]{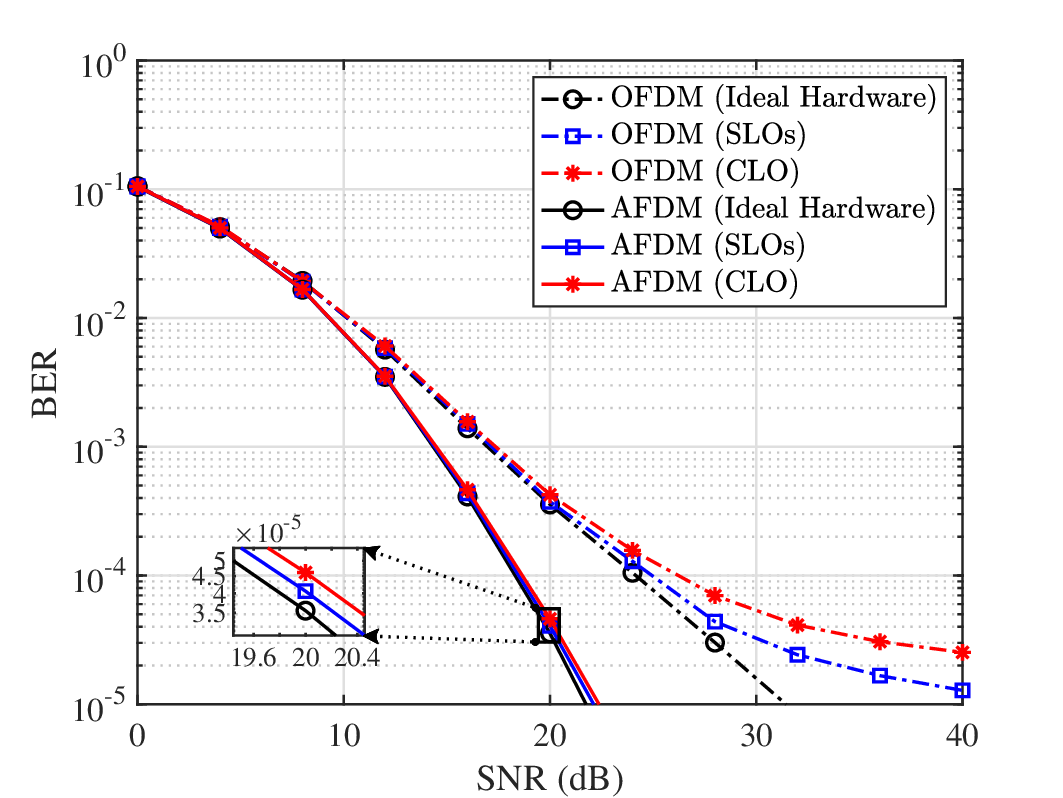}
\caption{BER performance of MIMO-AFDM/OFDM systems equipped with SLOs and CLO, corresponding to the effects of PN.}
\label{Figure4}
\end{figure}

We then compare the BER performance of MIMO-AFDM having ideal hardware and practical SLOs and CLO in Fig. \ref{Figure4}, where we use $\psi_\text{T}=\psi_\text{R}=10^{-17}$ \cite{bjornson2015massive}. Similar to Fig. \ref{Figure3}, it can be observed that AFDM systems attain similar BER performance for both ideal and imperfect LOs, while MIMO-OFDM suffering from SLOs is capable of achieving better BER than its CLO counterpart, since the PHN is averaged out in MIMO systems \cite{bjornson2015massive}. Specifically, at the BER of $3\times 10^{-5}$, an SNR gap of $15$ dB can be observed between the curves of MIMO-AFDM (CLO) and MIMO-OFDM (CLO). Furthermore, there is an SNR loss of $6$ dB and $9$ dB, when comparing MIMO-OFDM with CLO to its SLOs and ideal counterparts, respectively. The above observations confirm that AFDM is more resilient to phase noise than OFDM, regardless of the type of LOs. This is due to the full channel diversity achievable by AFDM, as well as owing to the chirp signalling's intrinsic resilience to phase changes.

\begin{figure}[b]
\centering
\includegraphics[width=0.9\linewidth]{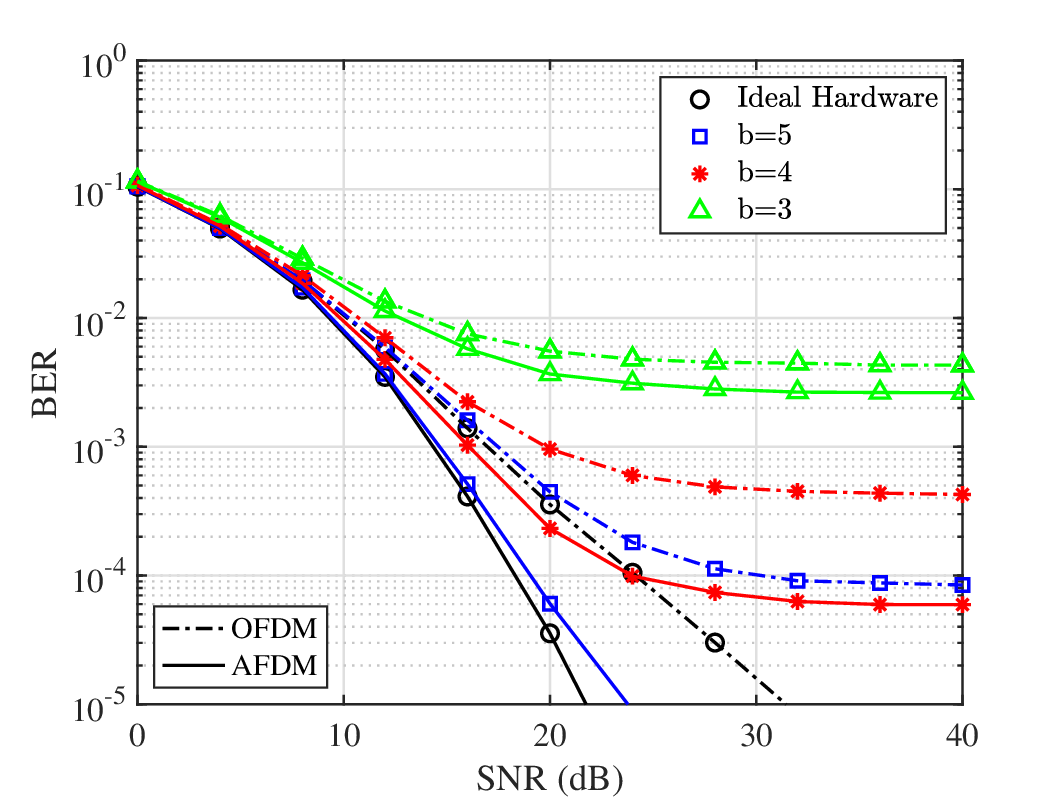}
\caption{BER performance of MIMO-AFDM/OFDM systems using low-resolution DACs with different numbers of quantization bits.}
\label{Figure6}
\end{figure}

In Fig. \ref{Figure6}, the BER performance of MIMO-AFDM using low-resolution DACs is characterized for $b=\{3,4,5\}$ quantization bits. Observe from Fig. \ref{Figure6} that low-resolution DACs yield a significant BER degradation; the lower the values of $b$, the higher the BER. Moreover, given $b=5$ and BER of $10^{-4}$, the MIMO-AFDM system only suffers about $1$ dB SNR loss compared to the ideal hardware case, while its OFDM counterpart exhibits about $8$ dB SNR erosion. This can be explained by the following details. The frequency-domain orthogonality of OFDM subcarriers may not be retained under substantial nonlinear distortions imposed by low-resolution DACs, implying that there is higher energy leakage among OFDM subcarriers compared to AFDM, yielding severe ICI and out-of-band emissions \cite{jacobsson2019linear}. By contrast, each AFDM chirp subcarrier spreads across the entire time-frequency domain. Hence, the uncorrelated quantization noise can be averaged with the aid of the DAFT process, implying that AFDM is capable of achieving better BER performance and higher SINR. Furthermore, the BERs of MIMO-AFDM and MIMO-OFDM systems relying on $b=\{3,4\}$ saturate, when the SNR exceeds $24$ dB and $32$ dB, respectively. This trend is attributable to the quantization noise and the scaling factor dominating the BER performance.

\begin{figure}[t]
\centering
\includegraphics[width=0.9\linewidth]{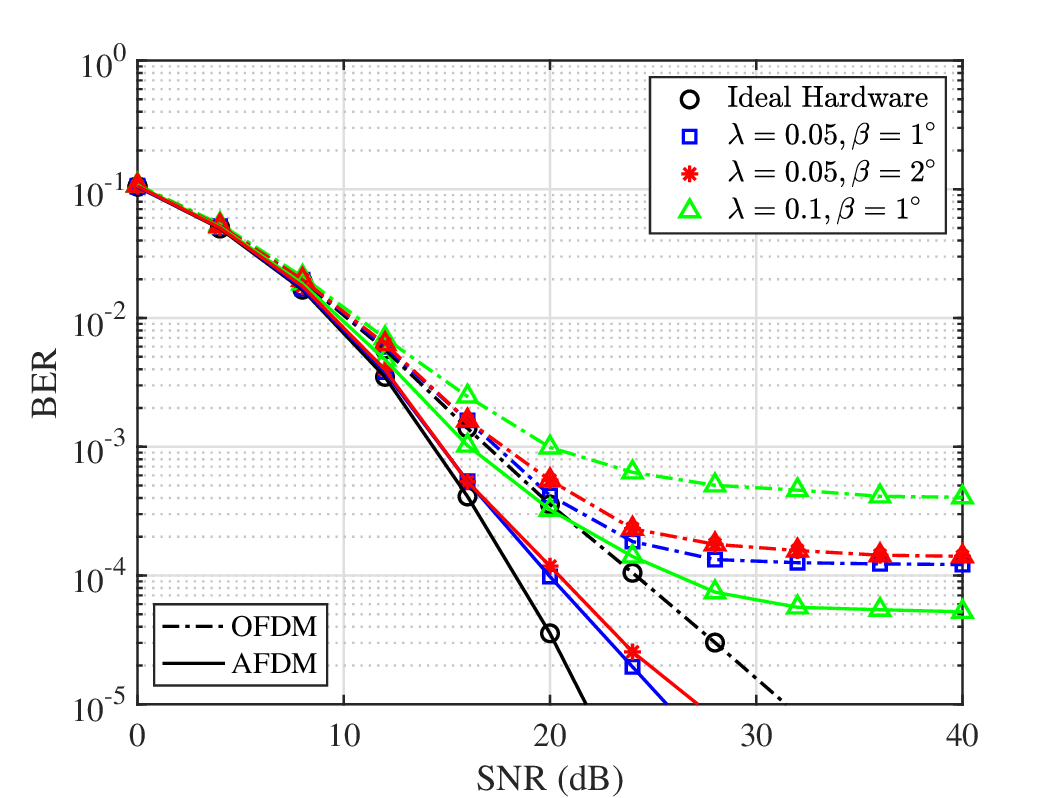}
\caption{BER performance of MIMO-AFDM and MIMO-OFDM systems under different IQI parameters.}
\label{Figure7}
\end{figure}

In Fig. \ref{Figure7}, we depict the BER performance of both ideal and IQI-infested MIMO-AFDM and MIMO-OFDM systems using LMMSE detection, where we harness ``$\lambda=0.05$, $\beta=1^\circ$'', ``$\lambda=0.05$, $\beta=2^\circ$'', and ``$\lambda=0.1$, $\beta=1^\circ$'' IQ parameter settings. As seen in Fig. \ref{Figure7}, among all the AFDM systems, the cases of ``$\lambda=0.05$, $\beta=2^\circ$'' and ``$\lambda=0.05$, $\beta=1^\circ$'' yield an SNR loss of about $6$ dB and $4$ dB over the ideal condition, respectively, while the BER curve of ``$\lambda=0.1$, $\beta=1^\circ$'' exhibits a floor at $5\times 10^{-5}$, when the SNR is higher than $32$ dB. This trend is driven by the dominance of the mirror interference term in \eqref{eq_HWI_IO}, whose power is comparable to that of the desired signal term. Furthermore, given the IQI parameters of ``$\lambda=0.05$, $\beta=1^\circ$'' and the BER of $2\times 10^{-4}$, the MIMO-AFDM system is capable of attaining an $8$ dB SNR gain compared to MIMO-OFDM. This is owing to ICI terms contaminating the mirror subcarriers, as shown in our analysis of Subsection \ref{Section 3-2-2} and \eqref{eq_HWI_IO}. By contrast, due to the intrinsic ICI resilience of chirp subcarriers and the spreading effect of DAFT modulation, AFDM achieves higher time-frequency diversity and, consequently, improved BER compared to OFDM using the same IQI parameters.

\begin{figure}[t]
\centering
\includegraphics[width=0.9\linewidth]{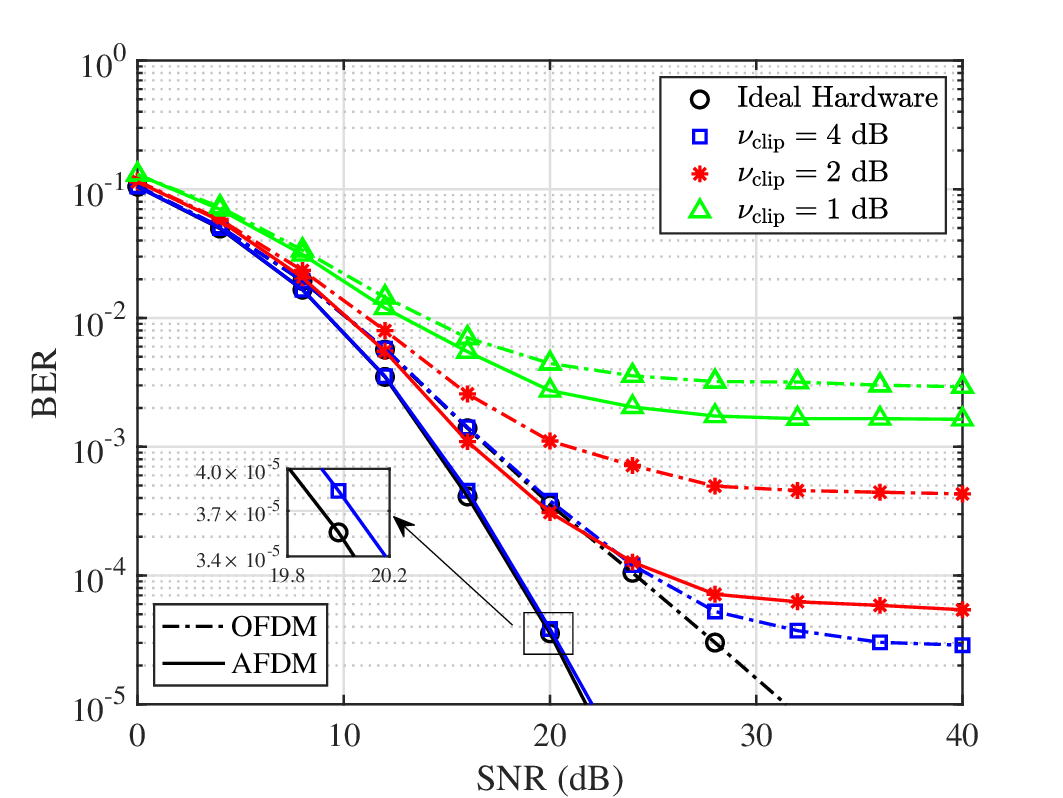}
\caption{BER versus SNR for MIMO-AFDM/OFDM systems with SEL-modelled NPA operating at different clipping levels.}
\label{Figure8}
\end{figure}

The BER performance of MIMO-AFDM/OFDM systems equipped with nonlinear PA operating at the clipping levels of $\nu_\text{clip}=\{4,2,1\}$ dB is shown in Fig. \ref{Figure8}. We infer the following observations from Fig. \ref{Figure8}. Firstly, at a given clipping level, MIMO-AFDM consistently achieves better BER than its MIMO-AFDM counterparts. This trend is attributable to the fact that AFDM is more resilient to ICI due to its chirp signalling and the spreading effect of DAFT. This is similar to the rationale behind Fig. \ref{Figure7}. Explicitly, the energy of NPA distortions spreads across different frequency bins, yielding in-band energy leakage, i.e., ICI of the spectral regrowth phenomenon \cite{fernandes2012analysis}. Moreover, the NPA-induced ICI behaves similarly to thermal noise, yielding a decision ``ambiguity'' in the constellation, which is similar to the PN-induced ICI. Secondly, the BER of MIMO-AFDM associated with $\nu_\text{clip}=4$ dB is nearly identical to that of MIMO-AFDM employing ideal hardware. This trend can be explained by the fact that the ideal PA can be regarded as having $\nu_\text{clip}=6$ dB, and AFDM is resilient to this minimal ICI imposed by a PA with $\nu_\text{clip}=4$ dB. By contrast, there is an $8$ dB SNR gap between the ideal and $\nu_\text{clip}=4$ dB scenarios employing MIMO-OFDM at a BER of $3\times 10^{-5}$. Additionally, except for the MIMO-AFDM scenario with $\nu_\text{clip}=4$ dB, at high SNRs, a constant BER is observed among the non-ideal PA cases in MIMO-AFDM/OFDM systems. This can be explained by the fact that the NPA not only generates $K_\text{PA}$, which reduces the power of the desired signal term in \eqref{eq_HWI_IO}, but also introduces DC interference and nonlinear-distortion terms.

\begin{figure}[b]
\centering
\includegraphics[width=0.9\linewidth]{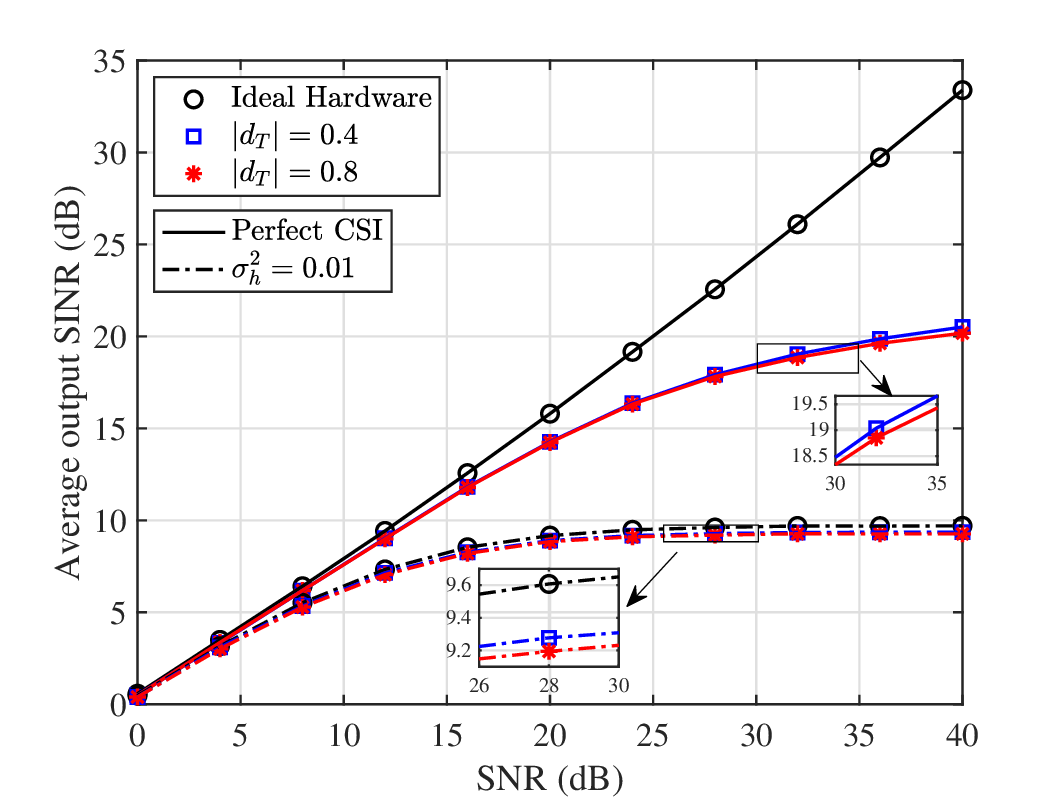}
\caption{Average output SINR of the LMMSE detector of MIMO-AFDM systems versus input SNR operating at different DCO factors and imperfect CSI.}
\label{Figure9}
\end{figure}

Fig. \ref{Figure9} shows the average output SINR of the LMMSE detector in MIMO-AFDM systems, where different DCO factors $d_\text{T}$ as well as both perfect and imperfect CSI ($\sigma_h^2=0.01$) conditions are considered, while we set $M=J=2$. From Fig. \ref{Figure9}, we have the following observations. Firstly, given the input SNR and channel estimation error $\sigma_h^2$, higher values of $|d_\text{T}|$ lead to lower average output SINR. It stems from the DC interference term of \eqref{eq_HWI_IO} that becomes more pronounced under this scenario. Secondly, when perfect CSI is achieved, DCO can significantly reduce the output SINR compared to the ideal hardware scenario. Specifically, the average output SINR operating in perfect CSI scenarios can be reduced by up to $39.6$\% compared to the $d_\text{T}=0.4$ case. This observation can be explained by the fact that DCO introduces the DC interference term in \eqref{eq_HWI_IO}, yielding high values of interference in the SINR of \eqref{eq_SINR_MMSE}. However, the output SINRs of the $d_\text{T}=0.4$ and $d_\text{T}=0.8$ paradigms are nearly identical, regardless of whether the CSI is perfect. Since other HWIs are not considered in this figure, it is challenging to further quantify the amount of DCO-induced interference by using higher values of $d_\text{T}$, although we have already exploited high DCO factors. As a matter of fact, the overall interference is mainly imposed by doubly-selective channels in this scenario. This observation is consistent with \cite{yih2009analysis}. Moreover, the average output SINRs become saturated under the scenario of imperfect CSI and moderate to high SNR, since the channel estimation error dominates them. Consequently, DCO only imposes approximately $0.4$ dB of output SINR reduction at an input SNR of $28$ dB, when $\sigma_h^2=0.01$.

\begin{figure}[b]
\centering
\includegraphics[width=0.9\linewidth]{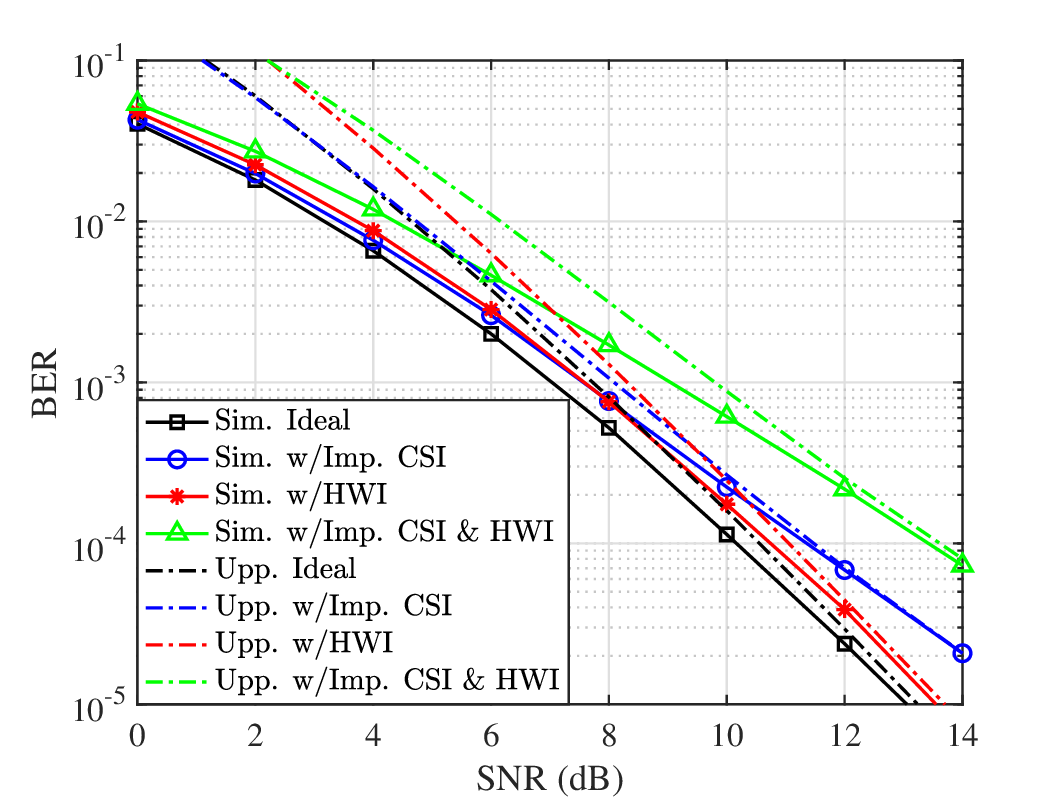}
\caption{BER performance of the ML detector and the upper bounds of MIMO-AFDM systems based on \eqref{eq_ABER}, where both imperfect CSI (``Imp. CSI'') and hardware imperfection (``HWI'') are considered.}
\label{Figure10}
\end{figure}

Then, we characterize the BER performance of the ML detector and the theoretical BER upper bounds of \eqref{eq_ABER} in Fig. \ref{Figure10}, where both imperfect CSI (Imp. CSI) using $\sigma_h^2=0.02$ and HWI scenarios are investigated. The HWI parameters are shown as Scheme 1 in Table \ref{table3}. Moreover, $N=8$ subcarriers and BPSK modulation are employed, and we harness $M=1$, $J=2$, and two-path doubly-selective channels. Based on the simulation results of Fig. \ref{Figure10}, we have the following observations. Firstly, the BER upper bound derived becomes tight at moderate to high SNRs, which validates the accuracy of our analytical results. Moreover, for a given BER of $10^{-4}$, the ``Imp. CSI'' and ``HWI'' cases suffer from approximately $1.5$ dB and $0.8$ dB lower SNRs than the ideal counterpart, while the ``Imp. CSI \& HWI'' scenario yields $3.8$ dB SNR loss. Furthermore, it can be observed that the ``Ideal'' and ``HWI'' BER curves have the same slopes, implying that the hardware-impaired AFDM systems can also achieve full diversity, as analyzed in Subsection \ref{Section 5-1}.

\begin{table}[t]
\vspace{-1em}
\footnotesize
\begin{center}
\caption{HWI Parameters in Simulations}
\label{table3}
\begin{tabular}{l|c|c|c|c|c|c|c}
\hline
Settings & $|d_\text{T}|$ & $b$ & $\varphi_\text{CFO}$ & $\beta$ & $\lambda$ & $\nu_\text{clip}$ & PN \\
\hline
\hline
Scheme 1 & 0.02 & 5 & 0.04 & $1^\circ$ & 0.02 & 4 dB & SLO \\  
\hline
Scheme 2 & 0.04 & 5 & 0.04 & $1^\circ$ & 0.05 & 4 dB & CLO \\  
\hline
\end{tabular}
\end{center}
\vspace{-1em}
\end{table}

\begin{figure}[t]
\centering
\includegraphics[width=0.9\linewidth]{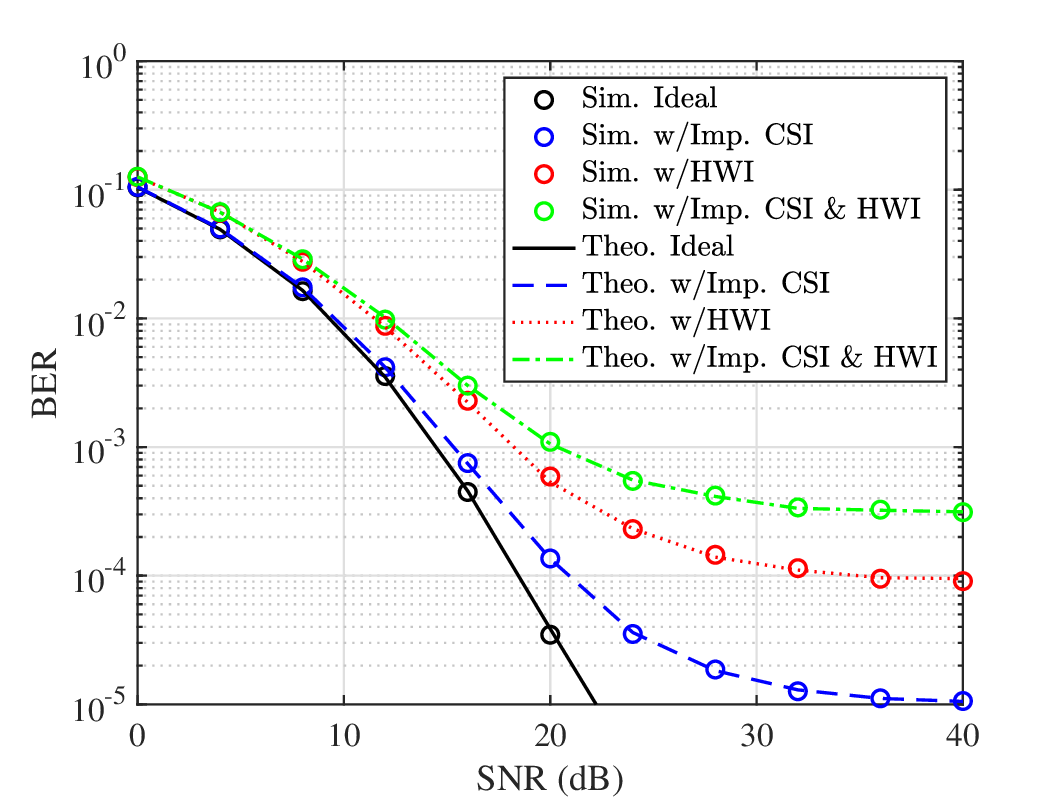}
\caption{BER performance of LMMSE detector and the theoretical BER based on \eqref{eq_Pe}, where both imperfect CSI (``Imp. CSI'') and hardware imperfection (``HWI'') are considered.}
\label{Figure11}
\end{figure}

Furthermore, in Fig. \ref{Figure11}, we provide the BER performance of the
LMMSE detector and compare it to our closed-form expression of \eqref{eq_Pe}. Similar to Fig. \ref{Figure10}, both imperfect CSI (Imp. CSI) with $\sigma_h^2=0.005$ and HWI scenarios are characterized. The HWI parameters are represented by Scheme 2 of Table \ref{table3}. From Fig. \ref{Figure11} we have the following observations. Firstly, among all the system settings, we observe a close match between the closed-form BER expression and the Monte-Carlo simulation results, even when both imperfect CSI and hardware are considered. This validates the accuracy of our analytical results. Secondly, for ``Imp. CSI'', ``HWI'' and ``Imp. CSI \& HWI'' cases, the BER remains constant at $10^{-5}$, $9\times 10^{-5}$ and $3.2\times 10^{-4}$ when the SNRs are higher than $36$ dB, $36$ dB and $32$ dB, respectively, since the channel estimation error and/or interference imposed by HWIs dominate the BER performance, even if the values of SNRs escalates. Moreover, at the given BER of $10^{-4}$, the ``HWI'' scheme attains approximately $15$ dB and $18$ dB lower SNR than the ``Imp. CSI'' and ideal settings, respectively.

\begin{figure}[b]
\centering
\includegraphics[width=0.9\linewidth]{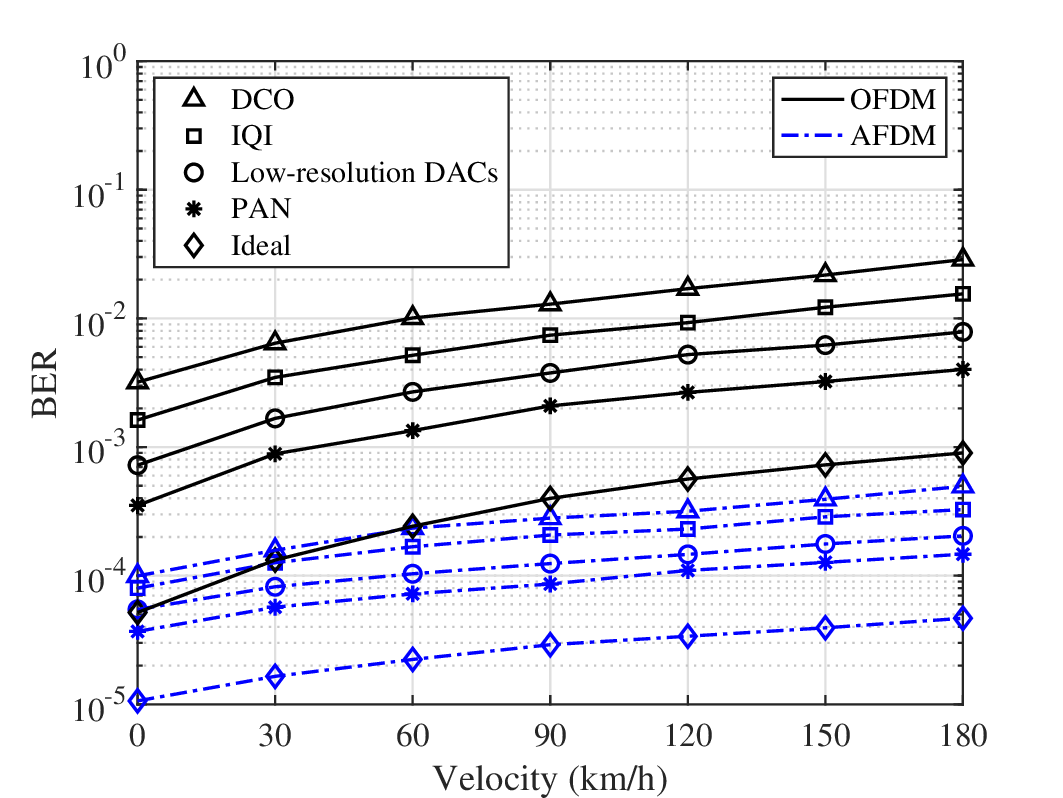}
\caption{BER comparison of AFDM and OFDM with different velocity values, whereby perfect CSI is assumed, where the HWI parameters are shown in Scheme 1 of Table \ref{table3}.}
\label{Figure12}
\end{figure}

\begin{figure}[t]
\centering
\subfigure[]{\label{Figure12-1}\includegraphics[width=0.9\linewidth]{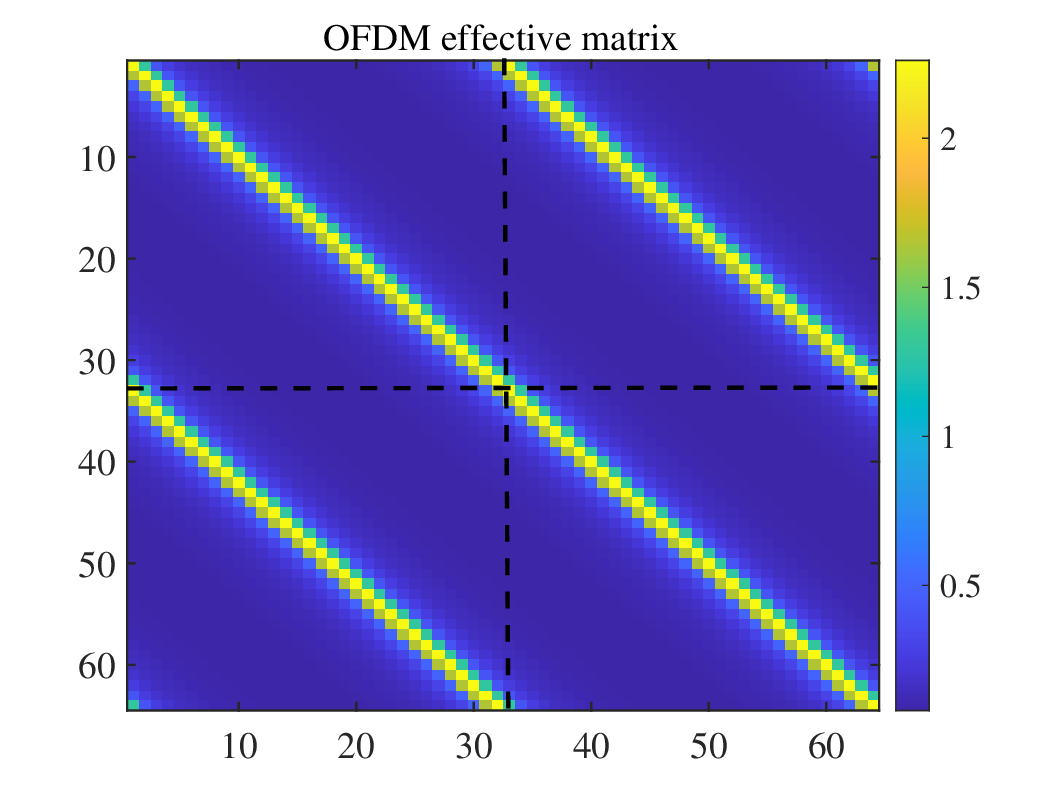}}
\subfigure[]{\label{Figure12-2}\includegraphics[width=0.9\linewidth]{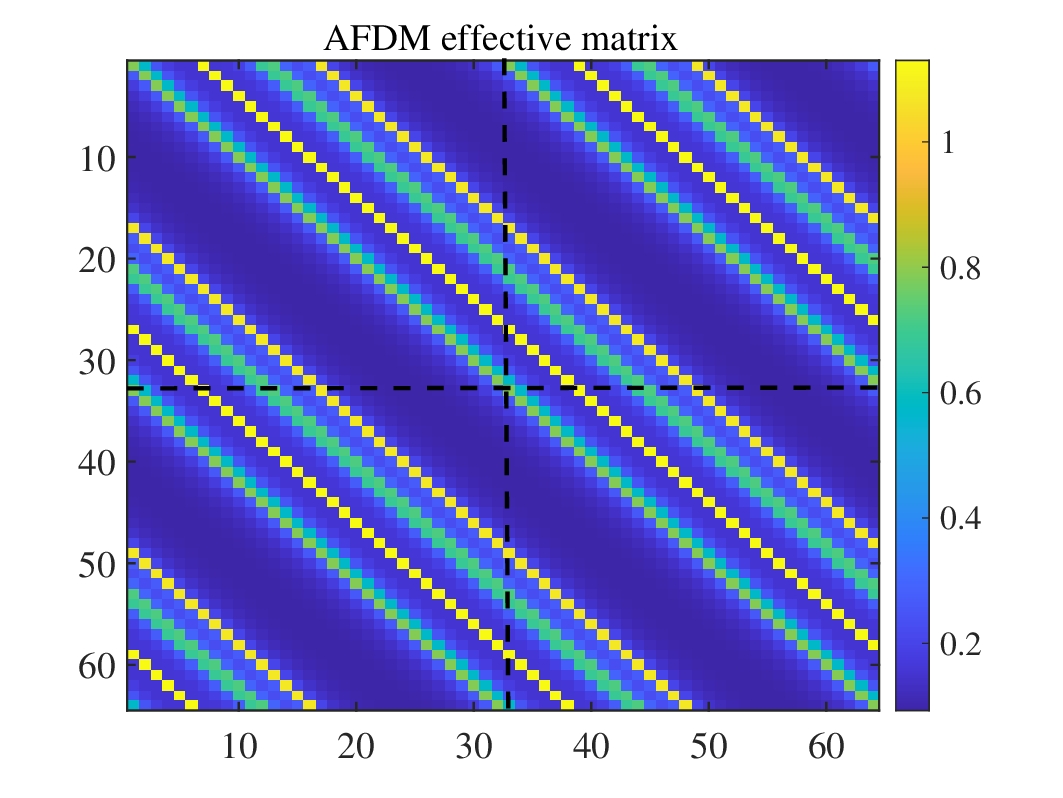}}
\caption{{Illustration of the HWI-infested effective channel matrices of \subref{Figure12-1} OFDM in frequency-domain and \subref{Figure12-2} AFDM in DAFT-domain based on \eqref{eq_MIMO_channel}, by using the HWI parameters in Scheme 1 of Table \ref{table3}.}}
\label{Figure12_1}
\end{figure}

In Fig. \ref{Figure12}, the BER performance of MIMO-AFDM/OFDM systems vs. velocity is studied, where the LMMSE detector and additive HWIs, i.e., DCO, IQI, low-resolution DACs, and NPA are considered. We employ $N=32$ subcarriers for a system operating at the SNR of $20$ dB. The HWI parameters are represented by Scheme 1 of Table \ref{table3}. From Fig. \ref{Figure12} we can have the following observations. Firstly, as expected, the higher the velocity, the worse the BER performance for both AFDM and OFDM. Moreover, MIMO-AFDM outperforms MIMO-OFDM even in the face of additive HWIs in the static or low-mobility scenarios, whereby additional ICI is imposed. This trend is also observed for our input-output relationship of \eqref{eq_HWI_IO}: when the system is free of PN and CFO, i.e., $\bm{\Phi}_{\text{T}}=\bm{I}_{NM}$, $\bm{\Phi}_{\text{R}}=\bm{P}=\bm{I}_{NJ}$, namely when the nonlinear distortion terms of mirror interference, DC interference and nonlinear-distortion interference still exist. This is aligned with our analysis and observations from Figs. \ref{Figure6}-\ref{Figure8}. {Furthermore, it can be observed that MIMO-AFDM outperforms MIMO-OFDM under ideal hardware scenarios, even in static conditions. This trend stems from the fact that uncoded AFDM is capable of extracting multipath diversity from delay spread, even in the absence of Doppler spread. By contrast, uncoded OFDM is unable to exploit any multipath diversity, unless frequency-domain spreading is used across the subcarriers, for example.} More importantly, we can see from all the simulation figures that OFDM is more sensitive to the ICI generated by both doubly selective channels and HWIs. By contrast, AFDM achieves better BER performance, validating our analytical results in Subsection \ref{Section 5-1}. Explicitly, AFDM can always achieve full diversity and ICI resilience.

{Next, we compare the effective channel matrices of MIMO-OFDM (in frequency domain) and MIMO-AFDM (in DAFT-domain) under HWIs in Fig. \ref{Figure12_1}. Specifically, we adopt $M=J=2$ TAs/RAs and $N=32$ subcarriers, a doubly selective channel having $P=4$ paths, and the maximum normalized Doppler shift is set to $k_\text{max}=1$. The corresponding HWI parameters are provided in Scheme 1 of Table \ref{table3}. For the AFDM channel matrix, one can see that the four dominant diagonal and sub-diagonal lines (corresponding to the four paths) do not overlap with each other, thus helping to achieve the full channel diversity. By contrast, in the OFDM channel matrix, only one dominant diagonal line can be observed, indicating that full channel diversity cannot be achieved. The above observations are consistent with our theoretical analysis in Subsection \ref{Section 5-1}.}

\begin{figure}[b]
\centering
\includegraphics[width=0.9\linewidth]{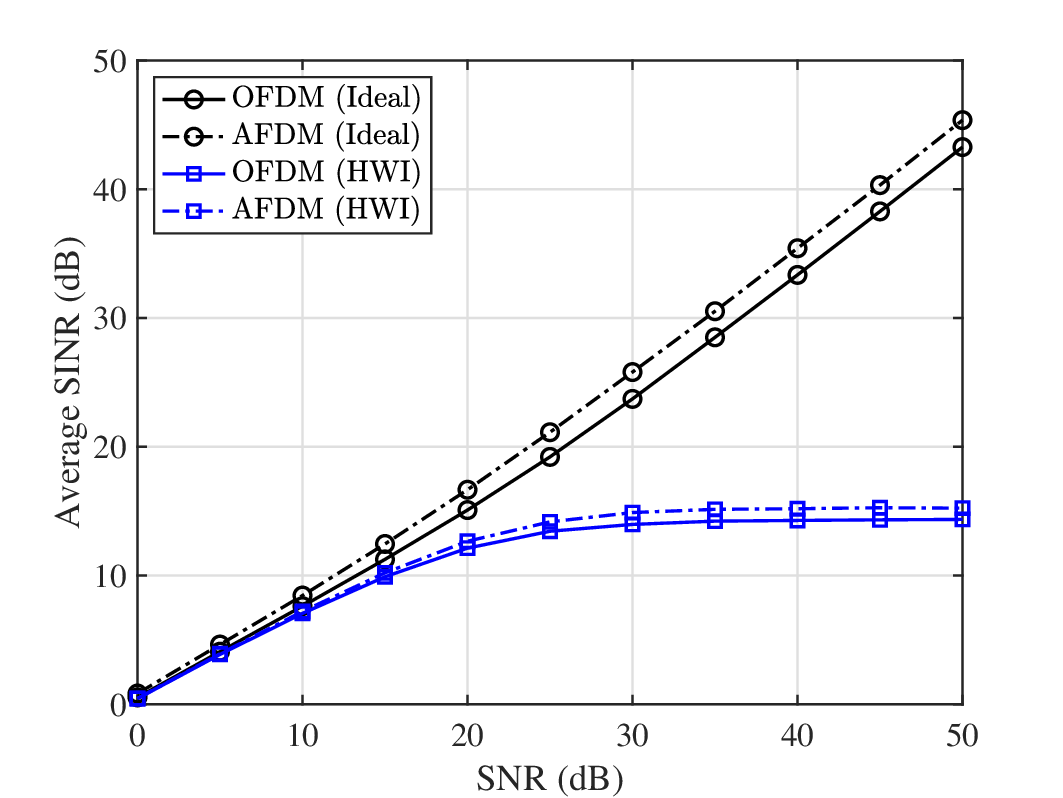}
\caption{{Average output SINR of the LMMSE detector of MIMO-OFDM and MIMO-AFDM systems based on \eqref{eq_SINR_MMSE}, where both ideal and imperfect hardware scenarios are considered, by using the HWI parameters in Scheme 1 of Table \ref{table3}.}}
\label{Figure13}
\end{figure}

{To further justify the superior performance of AFDM under HWIs, we compare the average output SINR of LMMSE detectors in Fig. \ref{Figure13}, calculated via \eqref{eq_SINR_MMSE}. Moreover, we set $N=16$ and $P=3$, while the remaining parameters are the same as those in Fig. \ref{Figure12_1}. It can be observed that AFDM achieves a higher average output SINR in both ``ideal'' and ``HWI'' cases. Furthermore, in the ``HWI'' scenario, the output SINRs of both AFDM and OFDM become saturated when the values of SNRs are higher than $30$ dB. Both of the above observations show that AFDM is more resilient to ICI in the face of doubly selective channels and HWIs than OFDM.}

\begin{figure}[t]
\centering
\includegraphics[width=0.9\linewidth]{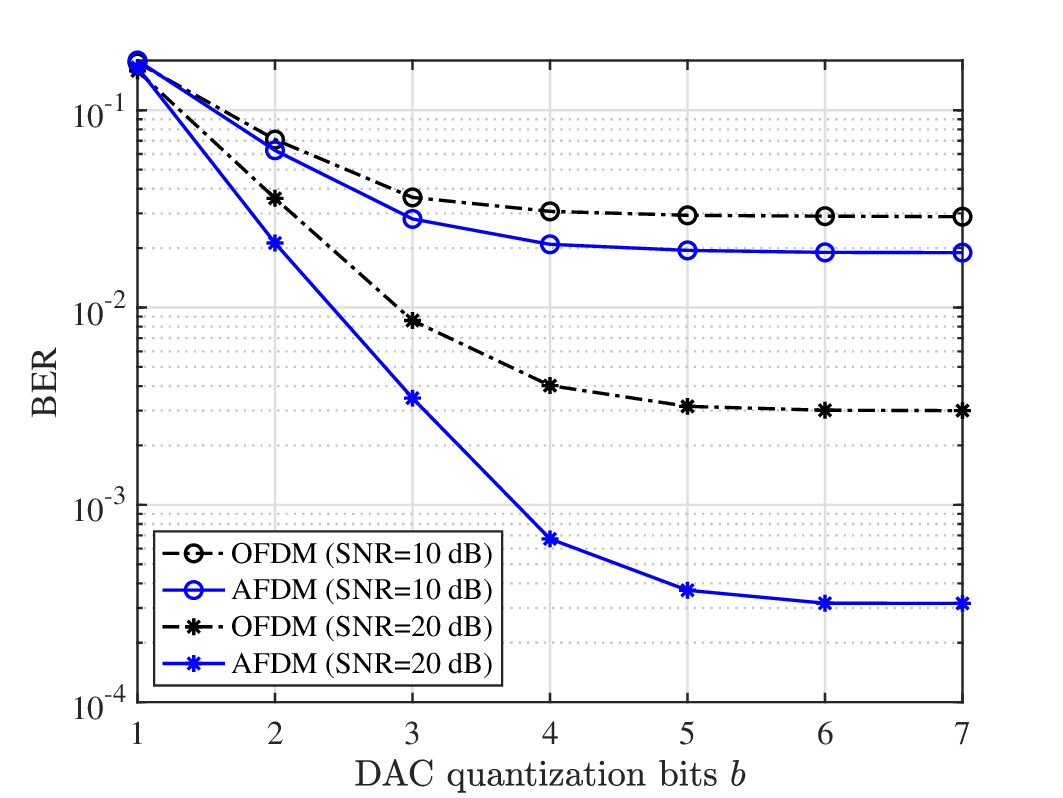}
\caption{{BER performance of both MIMO-OFDM and MIMO-AFDM versus different DAC quantization bits, where SNR$=\{10,20\}$ dB are considered.}}
\label{Figure14}
\end{figure}

\begin{figure}[b]
\centering
\includegraphics[width=0.9\linewidth]{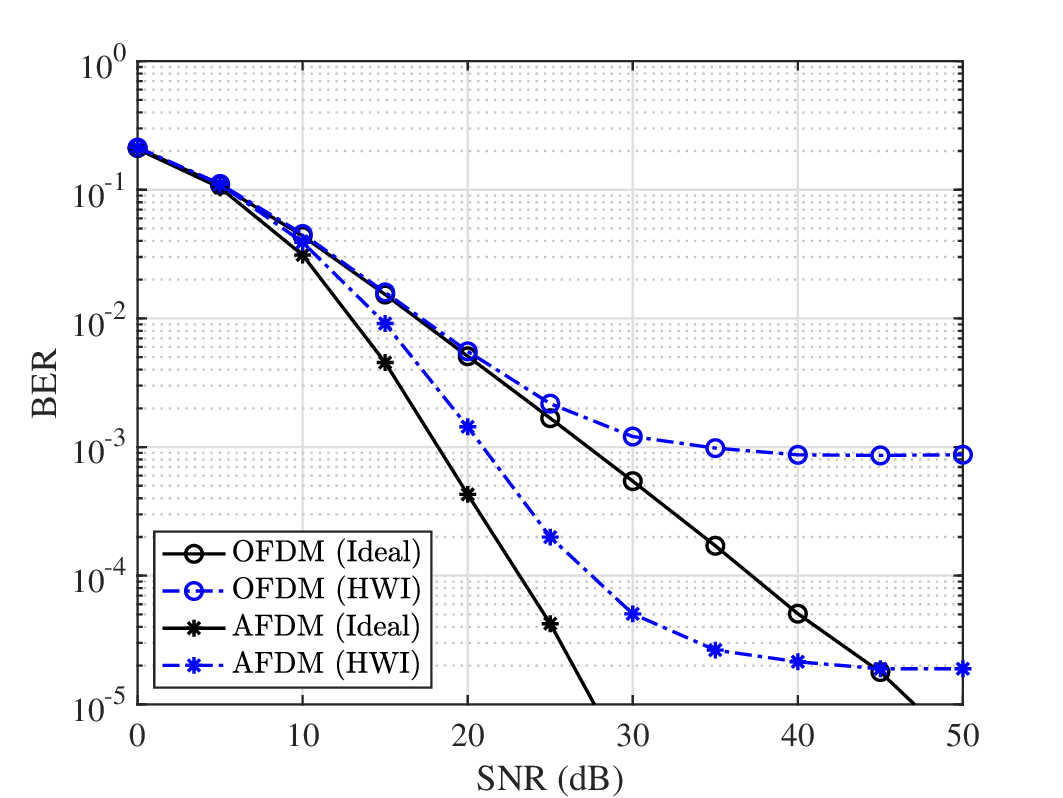}
\caption{{BER of SISO-OFDM and SISO-AFDM using LMMSE detectors via $9$-path EVA channels \cite{eva}, where both ideal and imperfect hardware scenarios are considered, by using the HWI parameters in Scheme 1 of Table \ref{table3}.}}
\label{Figure15}
\end{figure}

{In Fig. \ref{Figure14}, we investigate the impact of different numbers of quantization bits on the BER performance, where other parameters are the same as those in Fig. \ref{Figure13}. From Fig. \ref{Figure14}, we have the following observations. Firstly, the BER floors emerge at $b=4$ and $b=5$ under SNRs of $10$ dB and $20$ dB, respectively, regardless of the waveform. This is because the DAC scaling factor is close to zero, and the quantization noise can be omitted in the above DAC cases. Therefore, we may use fewer DAC quantization bits under low-SNR scenarios, since the performance gain of using more quantization bits is marginal. Moreover, the BER performance of AFDM is better than that of OFDM at a given SNR, indicating that AFDM is more resilient to ICI introduced by low-resolution DACs. When the number of DAC quantization bits is fixed, higher SNRs may introduce a more pronounced BER gap between OFDM and AFDM, since AFDM can attain higher diversity order in the face of HWI, which confirms our observations in Fig. \ref{Figure12_1}.}

{In Fig. \ref{Figure15}, we compare the BER performance of SISO-OFDM and SISO-AFDM using LMMSE detectors with $N=32$ subcarriers. We adopt the HWI parameters shown in Scheme 1 of Table \ref{table3}, and the Extended Vehicular A (EVA) channel \cite{eva} with $P=9$ paths. As predicted, when ideal hardware is considered, a $14$ dB SNR gap is observed at the BER of $10^{-4}$. Moreover, under the HWI scenario and when SNR is higher than $40$ dB, the BER curves of both SISO-OFDM and SISO-AFDM get saturated at about $9\times 10^{-3}$ and $2\times 10^{-5}$, respectively. This trend is due to the HWI-induced interference. The above-mentioned observations further validate the superior performance of AFDM systems under realistic channel models.}

\section{Conclusions}\label{Section 7}
This work has studied the performance of MIMO-AFDM systems under various types of HWIs, including both multiplicative and additive distortions. We first derived the corresponding input-output relationship in the face of additional HWI-induced interference terms. It was demonstrated that its diversity order remains full, as that of the AFDM systems with ideal hardware. Moreover, in small-scale MIMO-AFDM systems, we have derived the asymptotic BER upper bound corresponding to the ML detector. For large-scale MIMO-AFDM systems, an approximated BER associated with the LMMSE detector was presented. It was shown that the upper bound becomes tight as the SNR increases, and that the simulated BER is well aligned with the approximated BER. The impact of HWIs on the overall system performance has also been quantified. Simulation results illustrated that AFDM systems are resilient to both CFO and PN, while they are relatively sensitive to additive HWIs. Furthermore, it was found that MIMO-AFDM consistently outperforms MIMO-OFDM under identical HWIs and channel estimation error settings, thanks to its specific chirp subcarrier structure and full time-frequency diversity. 

{Observe from Figs. \ref{Figure6} and \ref{Figure14} that a higher DAC resolution reduces quantization distortion. Moreover, it can be seen from Figs. \ref{Figure4} and \ref{Figure8} that increased clipping threshold alleviates nonlinear distortion at the PA, and a lower phase-noise variance helps preserve phase coherence, respectively. In practical system design, parameters such as DAC resolution, clipping level, and phase-noise variance should be carefully selected to balance the performance metrics vs. the receiver complexity, hardware cost, and power consumption.}

{Future research should include synchronization, HWI mitigation, channel estimation, and data detection of hardware-impaired AFDM systems in the presence of forward error correction-aided mitigation. Specifically, besides conventional energy- and correlation-based schemes, machine learning solutions can also be harnessed for synchronization and TO compensation \cite{10136837}, as well as HWI mitigation \cite{9605580}. Since the HWI-infested DAFT-domain channel matrix still exhibits a sparse structure, we may utilize the family of compressed sensing algorithms, such as message passing and sparse Bayesian learning, to carry out channel estimation and data detection.}

{The impact of fractional delay on AFDM systems has been investigated in \cite{mirabella2026}. When the optimal chirp parameters $c_1$ and $c_2$ were determined for integer-delay channels, a BER performance loss was observed in fractional-delay scenarios. To date, a comprehensive performance analysis of AFDM under both fractional delay and Doppler is lacking. Additionally, similar to techniques such as cyclic delay diversity \cite{10845819}, space-time coding \cite{752125}, and cyclic delay-Doppler shift \cite{yin2026}, which achieve transmit diversity by introducing deterministic delays, time offset (TO) is also a constant at the transmitter side \cite{mirabella2026}. Nonetheless, any TO estimation error may degrade the receiver's BER performance. How to achieve a higher (or full) diversity order in fractional delay channels in the face of TO for AFDM systems is worthy of further investigation. In this case, machine learning may play a more effective role than conventional model-based approaches, in view of the need for both channel estimation and data detection.}
\renewcommand{\refname}{References}
\mbox{} 
\bibliographystyle{IEEEtran}
\bibliography{AFDM_HWI}
\end{document}